\documentclass[final,5p,times,twocolumn]{elsarticle}

\usepackage{amssymb}

\usepackage{siunitx}
\usepackage{mhchem}
\usepackage{url}
\usepackage{threeparttable}
\usepackage{notoccite}
\usepackage{mathtools}
\usepackage{relsize}
\usepackage{xcolor}
\usepackage{chemfig}
\usepackage{tikz}
\usepackage{subdepth}
\usepackage{pbox}

\newcommand{\ecoli}{{\it E.~coli}}

\usepackage{tikz}
\newcommand*\circled[1]{\tikz[baseline=(char.base)]{
            \node[shape=circle,draw,inner sep=0.75pt] (char) {#1};}}

\journal{Colloids and Surfaces B}

\begin{document}

\begin{frontmatter}



\title{ {\it Escherichia coli} as a model active colloid: a practical introduction}


\author{Jana Schwarz-Linek$^*$, Jochen Arlt$^*$, Alys Jepson$^*$, Angela Dawson$^*$, Teun Vissers$^*$, Dario Miroli$^{*\dagger}$, \\Teuta Pilizota$^\dagger$, Vincent A. Martinez$^{* (\ddagger)}$ and Wilson C. K. Poon$^{* (\parallel)}$}

\address{$^*$SUPA and School of Physics \& Astronomy, The University of Edinburgh, James Clerk Maxwell Building, Peter Guthrie Tait Road, Edinburgh EH9 3FD, UK, and $^\dagger$School of Biological Sciences, Darwin Building, Max Born Crescent, Edinburgh EH9 3BF, UK. $^{(\ddagger)}$vincent.martinez@ed.ac.uk, $^{(\parallel)}$w.poon@ed.ac.uk}

\begin{abstract}
The flagellated bacterium {\it Escherichia coli} is increasingly used experimentally as a self-propelled swimmer. To obtain meaningful, quantitative results that are comparable between different laboratories, reproducible protocols are needed to control, `tune' and monitor the swimming behaviour of these motile cells. We critically review the knowledge needed to do so, explain methods for characterising the colloidal and motile properties of \ecoli\, cells, and propose a protocol for keeping them swimming at constant speed at finite bulk concentrations. In the process of establishing this protocol, we use motility as a high-throughput probe of aspects of cellular physiology via the coupling between swimming speed and the proton motive force. 
\end{abstract}


\end{frontmatter}




Some time ago, our lab wanted to culture motile bacteria as `model active colloids'. We obtained a strain of {\it Escherichia coli} with the full complement of motility genes and a culturing protocol from a local microbiologist. For some time, we thought we were experimenting with motile \ecoli, until one day we checked in the microscope. Few, if any, of the cells were swimming! So we set out to learn how to modify the standard protocol to optimise motility by collating literature, talking to other researchers and trial and error; we also implemented differential dynamic microscopy (DDM) to quantify motility.

This article reviews what we have learnt. Some of the material is previously known, but seldom critically discussed in one place. We have explained some basic bacterial bioenergetics and genetics, because  physical scientists can use \ecoli\, and collaborate with biologists more effectively if these topics are understood. Much of the materials is new, arising from using DDM to quantify motility. While we aim primarily at researchers working on active colloids~\cite{poon2013physics}, this article should also be useful to others studying motility biophysics \cite{BerryReview}. 

From the outset, we refer to various culture media (BMB, TB, LB) and protocols, and freely use terminology related to molecular biology (plasmid, gene names, etc.) and measurement techniques (OD, DDM, etc.). Readers should refer to Section~\ref{sec:culture} on matters of cell culture, Sections~\ref{sec:ascolloid} and \ref{sec:DDM} for methodology, and Section~\ref{sec:genetools} and \ref{app:genotype} for biological jargon. We also provide a table of symbols in \ref{app:symbols}.

\section{\textbf{\textit{E. coli}} as a model active colloid}

An `active colloid' \cite{poon2013physics} is a suspension of $\approx \SI{5}{\nano\meter}$-\SI{5}{\micro\meter} particles that consume `fuel' to propel themselves, such as motile bacteria or various synthetic `colloidal swimmers' \cite{Howse,Sen}. Active colloids are out of thermal equilibrium even without external driving. Unusual phenomena displayed by such systems, such as `negative viscosity increment' \cite{AransonViscosity} and `rectification'  \cite{Galajda}, pose a `grand challenge' to statistical mechanics \cite{CatesReview2012,Joanny2014}. The self assembly of active colloids, possibly mixed with passive particles, may provide routes to new `smart materials' \cite{PalacciCrystal}. 

Progress in this new area will be greatly facilitated by experimental data from `model systems' that can most `cleanly' confront theory and simulations. Historically, well-characterised model passive colloids have enabled progress at critical points, from sedimentation equilibrium \cite{PerrinAtoms,Pais} to hard-sphere crystallisation  \cite{PvM} and sticky-sphere glass transitions \cite{Pham}.
The minimal requirements for a `model passive colloid' include reproducible synthesis/preparation, known particle size/shape distributions, quantifiable and `tuneable' interparticle interactions and accurate particle volume fractions~\cite{Puseychapter,PoonVarenna,Poonchapter,PoonVolfrac}. For model active colloids, we also need propulsion mechanisms that are understood and `tuneable', and knowledge of motility-specific interparticle interactions, e.g., via hydrodynamics. By these criteria, there is as yet no ideal model active colloid.

{\it Escherichia coli} is widely used in active colloids research. Its self propulsion is understood in essence  \cite{Berg,LiaoPair,DrescherPuller}, while propulsion mechanisms in many synthetic swimmers are still  debated \cite{brown14,ebbens14}. However, for \ecoli\, to become a fully-fledged model, reproducibility, control and characterisation need to be improved. We address these issues in this article.

\begin{table}[t]
\caption{Average length and standard deviation of 30 cell bodies of \ecoli\, AB1157 grown in LB/TB at \SI{30}{\celsius}/\SI{37}{\celsius} to late-exponential (OD = 0.5) or stationary (16 hours growth) phase. Note that these cells have not been washed in BMB. Washing, however, makes little difference. }
\smallskip
\begin{center}
\begin{tabular}{| l | l | l | l | l |}
\hline
 & LB$_{\rm 30, stat}$ & TB$_{\rm 30, stat}$ & LB$_{\rm 37, stat}$ & TB$_{\rm 37, stat}$ \\
 \hline
 $l \;(\si{\micro\meter})$ & 1.7 & 1.5 & 1.6 & 1.7\\
 $\delta l\; (\si{\micro\meter})$ & 0.4 & 0.3 & 0.3 & 0.5\\
 \hline
 $\delta l/l$ & 0.24 & 0.2 & 0.19 & 0.29\\
 \hline\hline
 & LB$_{\rm 30, 0.5}$ & TB$_{\rm 30, 0.5}$ & LB$_{\rm 37, 0.5}$ & TB$_{\rm 37, 0.5}$\\
 \hline
$l \;(\si{\micro\meter})$ & 	4.4 & 2.3 & 4.4 & 2.4 \\
$\delta l\; (\si{\micro\meter})$ & 1.2 & 0.6 & 1.0 & 0.6\\
\hline
$\delta l/l$ & 0.27 & 0.26 & 0.23 & 0.25 \\
\hline
 \end{tabular}
 \end{center}
 \label{tab:length}
 \end{table}
 
\section{\textbf{\textit{E. coli}} as a bacterium}
\label{}

{\it Escherichia coli} \cite{Escherich} is the best understood living organism on earth today. It is often {\it the} model of choice for understanding molecular biological processes: `Tout ce qui est vrai pour le Colibacille est vrai pour l'\'el\'ephant.'~\cite{Jacob}  

The cytoplasm of \ecoli\, is enclosed by two lipid bilayer membranes. Between them is a peptidoglycan `cell wall' (or `periplasmic space'), a network of poly-sugars (`glycans') linked by short peptides. Within the species known as \ecoli\, there are many genetic variants, or `strains' (so that `bacterial species' is a problematic concept \cite{species}). The genome of strain MG1655 was first to be sequenced \cite{coligenome}; today, the genome of 2085 strains are available in GenBank \cite{GenBank}, although  $ \gtrsim 30\%$ of the genes remain of unknown function. The best single source of information on \ecoli\, (and its close relative  {\it Salmonella}) is a two-volume `bible' \cite{EcoSalBook} now updated digitally~\cite{EcoSalPlus}.

Each cell carries multiple helical flagella \cite{Calladine1976,Calladine1978} powered by rotary motors \cite{BerryReview,Berg} embedded in the membranes and cell wall \cite{McNab1996}. When all flagella rotate counterclockwise (CCW) (viewed from behind), they bundle to propel the cell forward. When a motor reverses to clockwise (CW), its flagellum unbundles and the cell tumbles. Wild-type (WT) cells run ($\approx \SI{1}{\second}$), tumble ($\approx \SI{0.1}{\second}$), and run, the latter in a direction that is more or less uncorrelated with the run before the tumble \cite{BergBrown}. When successive sensings of the environment return an increasing  concentration of an `attractant', intracellular signals reduce the rate of CCW $\rightarrow$ CW $\rightarrow$ CCW switching (i.e. the tumbling rate). This biased random walk up an attractant gradient constitutes chemotaxis.

A number of \ecoli\, strains used for active colloids and related research are listed in Table~\ref{tab:strains}, \ref{app:strains}. Many of these were derived from K-12, a `debilitated' laboratory strain \cite{Hobman,Hayashi2001} that does not normally infect humans. It is deemed safe for laboratory use and classified as a category~1 biohazard \cite{CDC}. Mutants in which one or more key chemotaxis genes have been deleted, e.g. $\mathit{\Delta}${\it cheY}, lose the ability to tumble. These `smooth swimmers' simply swim forward, with a persistence time set by their inverse rotational diffusivity $D_{\rm r}^{-1} = 2\pi\eta L\sigma^2/k_B T \approx \SI{5}{\second}$, where we have modelled a cell body with its flagella bundle as an ellipsoid of major and minor axes $L \approx \SI{10}{\micro\meter}$ and $\sigma \approx \SI{1}{\micro\meter}$.

\section{\textbf{\textit{E. coli}} as a colloid}
\label{sec:ascolloid}

The cell body of an \ecoli\, bacterium is a net negatively charged spherocylinder (diameter $\sigma$, pole-to-pole length $l$).\footnote{For physicochemical `vital statistics' of \ecoli, see \cite{KeyNumbers2010,CyberCell,bionumbers}}  For mostly unknown reasons \cite{Young2007,NelsonSegregation}, bacterial shapes are highly conserved:\footnote{It is possible to turn bacteria into silica colloids \cite{Nomura2010}, opening up a novel route to synthetic colloids with many shapes \cite{Young2007}.} shape mutations tend to be lethal.  In this section, we review the single-particle `colloidal' properties of \ecoli\, and how to estimate cell concentrations.

\begin{figure}
\begin{center}
\includegraphics[width=0.35\textwidth]{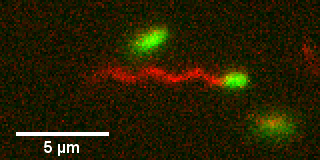}
\caption{Visualisation of both body and flagella bundle of swimming \ecoli\, from a real-time movie. Flagella are fluorescently labelled using Alexa546 and the body with a plasmid expressing GFP (AD1 pHC60, Table~\ref{tab:strains}). Using epi-fluorescence microscopy these can be imaged into separate channels at high enough frame rates to avoid motion blurr. Image part of a 100 fps movie acquired using an Orca Flash 4.0 camera on a Nikon Ti microscope with PA60x/1.4 OIL objective using a Cairn Optosplit II.}
\label{fig:flagella}
\end{center}
\end{figure}

\subsection{Shape and size}

\subsubsection{Microscopy}

The shape and size of \ecoli\, cell bodies can be obtained using microscopy and scattering (compared in \cite{Latimer1979}). Measuring cell bodies using phase-contrast microscopy of strain AB1157 harvested using our standard protocol and washed into BMB gave $\langle l \rangle =  2.4 \pm 0.6 \si{\micro\meter}$, where the uncertainty is a polydispersity. At mid-cell, $\langle \sigma \rangle = 0.86 \pm 0.07 \si{\micro\meter}$, where the uncertainty represents measurement errors. Importantly, these `vital statistics' depend on strain, growth conditions and the time of harvest, Table~\ref{tab:length}. Time of harvest is important because bacteria grow in well defined stages: from rapid division in an exponential phase (cell number density $n$ grows exponentially with time), through a stationary phase ($n = $ constant), to a death phase ($n$ decreases with time). Our data show that stationary-phase cells tend to be less polydisperse in length than mid-exponential cells, as found previously~\cite{Bernander1995}. There may also be some dependence of $\sigma$ on $l$~\cite{Trueba1980}. 

It is often desirable to visualise the flagella bundle, because it renders the cell highly anisotropic and introduces strong steric effects \cite{Solomon2010,YodhRod}. If the cell body and flagella are labelled with the same dye \cite{Berg2000,BergTorque}, fluorescence from the former will dominate. To visualise the flagella alone, the {\it fliC} gene\footnote{This encodes flagellin, the protein units used to build up flagella filament.} can be mutated to facilitate binding to Alexa Fluor C5 maleimide (e.g., Alexa546) \cite{Zhang2010}. If cells carry GFP-plasmid pHC60 (Table~\ref{tab:strains}), then the cell body and flagella can visualised simultaneously, Fig.~\ref{fig:flagella}. Note that such mutants should always be checked for potential changes to their motility and other phenotypes. 

\subsubsection{Scattering}

Cell shape and size can also be studied by static light scattering (SLS) either averaged over cell populations~\cite{Koch1968,Latimer1972,Morris1974,Chen1978,Latimer1982,Katz2003} or individually in flow cytometry~\cite{Maltsev2013}. Given information on the refractive index \cite{Bateman1965}, models can fitted to scattering data. Typically, an \ecoli\, cell with cytoplasm surrounded by a cell wall is modelled as a core-shell ellipsoid~\cite{Koch1968,Latimer1975,Barber1979a}, giving interference fringes that survive polydispersity in $l$ because there is very little variation in either $\sigma$ or periplasm thickness~\cite{Latimer1972,Barber1979b}. How optical inhomogeneities due to the chromosome  affect dynamic light scattering, which can also give information on size and shape, has been studied~\cite{Berne1974}. Polarised scattering  gives extra information \cite{Bronk1995,Bronk1997}.

\subsection{Sedimentation}

The (mass) density of \ecoli\, cells, $1.08 \lesssim \rho \lesssim \SI{1.16}{\gram\per\cubic\centi\metre}$, depends on media and conditions \cite{Koch1995a,Koch1995b,godin2007}. The density difference with the solvent, $\Delta \rho \lesssim \SI{0.1}{\gram\per\cubic\centi\metre}$, causes cells to sediment. The distribution of particles in a suspension in sedimentation equilibrium defines the colloidal length scale~\cite{Poonchapter}. In this dynamic equilibrium, concentration-driven diffusive flux from the  bottom is balanced by a sedimentation flux from the top, so that when the suspension is dilute enough to neglect interparticle interactions, the number density profile with height is
\begin{equation}
n(z) = n(0) e^{-z/\ell_g},
\end{equation}
where the gravitational height $\ell_g$ is given by
\begin{equation}
\ell_g = \frac{D_0}{v_s}. \label{eq:sedheight}
\end{equation}
Here, $D_0$ is the free-particle diffusivity, and $v_s$ is the single-particle sedimentation speed. A colloid is suspendable against gravity,  so that $\ell_g$ has to be larger than its characteristic size.

The friction coefficient for motion parallel to the long axis of a prolate ellipsoid with major and minor axes $l$ and $\sigma$ is \cite{BergRandom} 
\begin{equation}
\xi_\parallel = \frac{2\pi \eta l}{\ln\left(\frac{2l}{\sigma}\right) - \frac{1}{2}},
\end{equation}
where $\eta$ is the solvent viscosity. For \ecoli\, in typical motility experiments, $l/\sigma \approx 2$ to 3, so that $\xi_\parallel \approx 2\pi\eta l$, from which
\begin{equation}
v_s \approx \frac{\Delta\rho g \sigma^2}{12\eta}. \label{eq:sedspeed}
\end{equation}
Translational diffusion along the long axis is governed by
\begin{equation}
D_\parallel = \frac{k_B T}{\xi_\parallel},
\end{equation}
so that the gravitational height of a non-motile cell is
\begin{equation}
\ell_g^{\rm (coli)} \approx \frac{6k_B T}{\pi \sigma^2l\Delta \rho g} \approx \SI{4}{\micro\meter} \approx 2l.
\end{equation}
Thus, the cell body of non-motile \ecoli\, is just colloidal. Motile \ecoli\, should have significantly higher $\ell_g$ \cite{Bocquet}; and motility coupled with growth can display rich phenomena \cite{Freeman2008}.  

\subsection{Interactions: stability and surface adhesion}

\begin{figure}[t]
\begin{center}
\vspace*{-.6cm}
\includegraphics[width=0.42\textwidth]{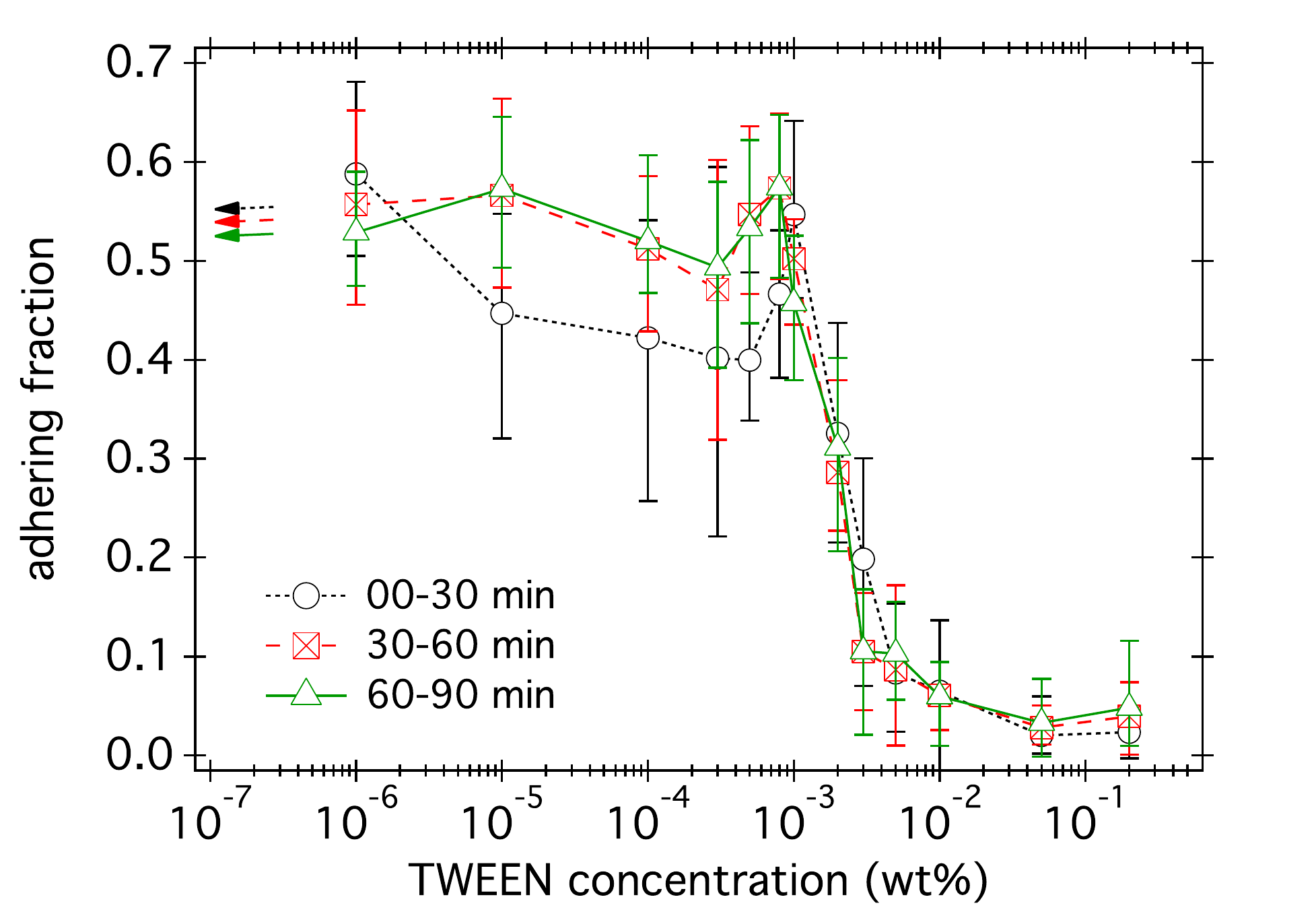}
\caption{Fraction of bacteria sticking to the lower capillary wall in a $\approx4$~cell width layer at different concentrations of TWEEN 20 surfactant in BMB with \SI{0.72}{\milli\mbox{M}} glucose for WT AB1157 at $\approx 6\times10^7$~cells/ml. Points are averages over $0-30$ minutes, $30-60$ minutes and $60-90$ minutes after loading. Arrows: adhering fractions with no surfactant. \label{fig:sticking} }
\end{center}
\end{figure}

The surface properties of \ecoli\, \cite{Rideal1956,Liu2013,Kaper2004} are strain dependent \cite{BosSurface}. 
Interpreting electrophoretic mobilities ($\mu$) is non-trivial because surface macromolecules \cite{Liu2013,BosSurface,Kaper2004} and the periplasm are ion-permeable. Even with a `soft' electrokinetic theory allowing for this \cite{Lyklema1997,Elimelech2005}, relating $\mu$ to zeta potentials and  surface structures is fraught. 

Nevertheless, \ecoli\,  definitely carries a net negative charge under physiological conditions ($-3\lesssim \mu \lesssim -2 \si{\micro\meter\centi\meter\per\volt\per\second}$) largely due to ionised carboxylate and phosphate groups \cite{BosSurface,Rideal1956}. This charge is normally sufficient to confer colloidal stability in growth media or in BMB.\footnote{The screening length in BMB is $\lesssim \SI{1}{\nano\meter}$.}  However, a strain that has behaved stably for many months may suddenly start aggregating due to `phase variation' in the expression of the surface protein Ag43 (antigen 43) \cite{Ag43Review}. Each generation, there is an $\mathbf{O}(10^{-3})$ probability that a cell switches between `on' and `off' states of Ag43 expression. `On' renders cells `sticky' and leads to `autoaggregation'.  The expression state is heritable and reversible. 

For active colloids work using \ecoli\, close to surfaces \cite{brown2015,TurnerFlagStain,Niu2005,Koumakis2013}, it is important to minimise adhesion, which is far from understood. Various non-ionic surfactants prevent \ecoli\, from sticking to untreated glass \cite{Berg2000,Koumakis2013}. Figure~\ref{fig:sticking} quantifies this effect for a popular surfactant used for this purpose.

We imaged AB1157 cells in BMB within an optical slice of $\approx 4$ cell width next to the bottom surface of an untreated glass capillary using a Mikrotron MC1362 camera and a Nikon Ti microscope with $60 \times$ phase contrast objective (PF60$\times$/0.7). Automated software for identifying \cite{Besseling2014} and tracking rods was used to analyse movies, classifying cells into `free' (diffusing or swimming) and `adhering' (stuck). The average fraction of adhering cells, extracted from many movies, dropped abruptly to $\gtrsim 0$ at a TWEEN 20 concentration of $\approx 10^{-2}$~wt.\%. Note that concentrations up to 0.2\% have been used previously \cite{Koumakis2013}.  Preliminary DDM data suggest that commercial TWEEN may contain impurities rapidly metabolisable by \ecoli, so that high enough concentrations of TWEEN may significantly alter swimming behaviour. 

\subsection{Cell concentration}

The definition of cell concentration and its determination are non-trivial. The cell body volume fraction, $\phi$, gives a better intuitive feel for the degree of `crowding' than the number density of cells, $n$.
The {\it effective} volume fraction, which takes into account the cell consists (body length $l$) {\it and} flagella bundle with total length $L \lesssim \SI{10}{\micro\meter}$ is $\phi_{\rm eff} \sim (L/l)^3 \phi \gtrsim 10^2 \phi$ if the cells are randomly oriented, so that they reach `overlap' at $\phi \lesssim 0.01$. 

The standard way to obtain $n$ is from the spectrophotometric optical density (OD) of cell suspensions, which is proportional to $n$ for $n \lesssim 10^9 \si{\mbox{cells}\per\milli\litre}$ (corresponding to $\phi_{\rm eff} \approx 0.2$).\footnote{The OD depends on the wavelength used; we use \SI{600}{\nano\meter} throughout.} Calibration is by `viable plate counting' (Section~\ref{sec:count}), and is found to depend on growth stage and other conditions.

\section{Characterising \textbf{\textit{E. coli}} motility}
\label{sec:DDM}

We turn next to characterising \ecoli\, as self-propelled particles. The average swimming speed, $\bar v$, and the fraction, $\beta$, of non-swimmers that only diffuse,\footnote{Note that the bulk diffusion is enhanced by the swimmers \cite{Jepson2013}. \label{note:enhance}} can be measured using a high-throughput technique, differential dynamic microscopy (DDM) \cite{CerbinoDDM1,WilsonDDM,MartinezDDM}. Application to swimming algae \cite{MartinezDDM} and passive magnetic rods \cite{Reufer} can be found elsewhere. Here we focus on practical aspects of using DDM to characterise motile \ecoli.

We take a low-magnification movie of a cell population, and calculate the squared modulus of the Fourier transform (F.T., reciprocal vector $\mathbf{q}$) of the difference of two images separated by $\tau$ in time, and averaged over starting times:
\begin{equation}
g(\mathbf{q},\tau) = \left\langle \left| \mbox{F.T.}\left[ {\cal I}(x,y;t+\tau) - {\cal I}(x,y;t) \right] \right|^2 \right\rangle. \label{eq:g}
\end{equation}
If the image intensity ${\cal I}(x,y;t)$ at pixel position $(x,y)$ at time $t$ is linearly related to the cell density at the corresponding sample position and time, then $g(\mathbf{q},\tau)$ is directly related to the density-density time correlation function, $f(\mathbf{q},\tau)$, of the system, with $q = |\mathbf{q}| \sim 2\pi(\mbox{image\,size})^{-1}$ to $\pi(\mbox{pixel\,size})^{-1}$ being experimentally accessible. Indeed, under appropriate conditions, 
\begin{equation}
g(q,\tau)=A(q)\left[1-f(q,\tau)\right]+B(q), \label{eq:DDM1}
\end{equation}
Sometimes called the normalised dynamic structure factor or intermediate scattering function (ISF), $f(q,\tau)$ is in principle (but not in practice~\cite{WilsonDDM}) also measurably by dynamic light scattering. We assume isotropy, so that $g$ and $f$ only depend on $q$. In Eq.~\ref{eq:DDM1}, $A(q)$ is related to the form and (static) structure factors of the bacterial suspension, and $B(q)$ is instrument noise. 

\begin{figure}
\begin{center}
\includegraphics[width=0.5\textwidth]{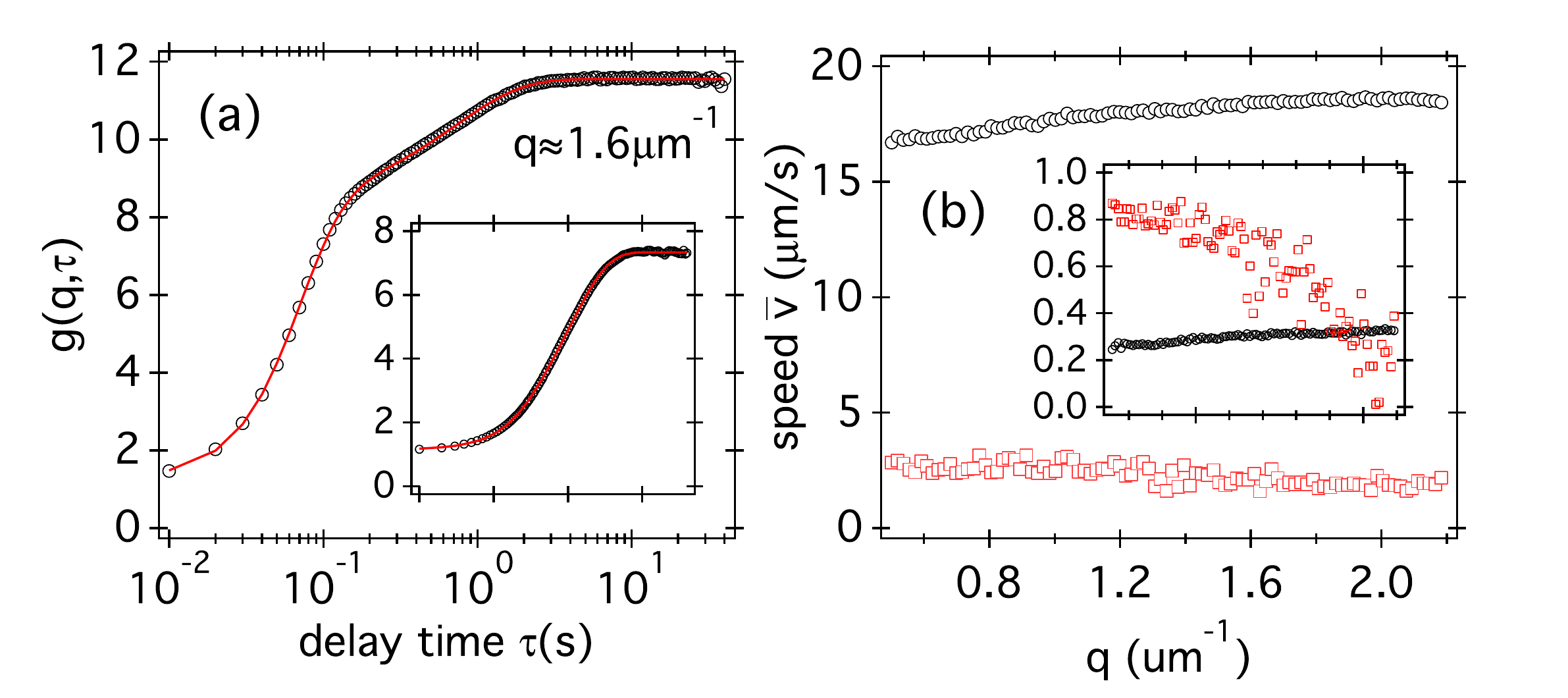}
\caption{(a) Fitting DDM data $g(q,\tau)$, Eq.~\ref{eq:g} and \ref{eq:DDM1}, using the ISF given by Eq.~\ref{eq:DDM2} at one $q$ for WT AB1157 at $5\times 10^8$cells/ml. Main plot: data for bacteria swimming fast enough for ballistic and diffusive motions to show up clearly as two processes, separated by  a point of inflection at $\tau \approx \SI{0.4}{\second}$. Inset: the bacteria have slowed down to the extent that swimming and diffusion are no longer well separated.  (b) Fitted speed $v$ (main plot) and non-motile fraction $\beta$ (inset) for the data shown in the main part of (a) (black) and in the inset of  (a) (red).}
\label{fig:easy_difficult_fitting}
\end{center}
\end{figure}

The ISF can be modelled once we know how cells move. For smooth-swimming \ecoli\, in 3D mixed with non-motile cells, 
\begin{equation}
f(q,\tau)=\beta e^{-q^2D\tau}+(1-\beta)e^{-q^2D\tau}\int_v P(v)\frac{\text{sin}(qv\tau)}{qv\tau}dv, \label{eq:DDM2}
\end{equation}
where $P(v)$ is the speed distribution of the swimmers, and $D$ is the free translational diffusion coefficient, which is mainly determined by the diffusivity of the non-motile cells (but see footnote~\ref{note:enhance}). We have ignored polydispersity in cell size (which induces a distribution in $D$) for simplicity. 

Self consistency demands that for straight swimmers mixed with diffusers, the dynamical quantities obtained by fitting  Eq.~\ref{eq:DDM2} to data, in particular $\bar v = \int_0^\infty vP(v) dv$, $\beta$ and $D$, should be $q$-independent. Residual $q$ dependence is nevertheless observed in practice, Fig.~\ref{fig:easy_difficult_fitting}(b), e.g. due to tumbling\footnote{Calculations of the effects of tumbling on the ISF exist \cite{Leonardo2013}.} \cite{MartinezDDM}; such $q$ dependence is the main source of experimental uncertainties in the fitted parameters from DDM quoted in this work. 
 
The average body rotation angular frequency, $\bar \Omega$, can be determined using another high-throughput technique, dark-field flicker microscopy (DFM) \cite{Martinez2014}. Essentially, the lowest peak in the Fourier transform of the time-dependent image intensity of a single cell in dark-field microscopy is $\Omega$. In the low-Reynolds-number limit, $\bar v /\bar \Omega$ depends solely on the geometry of the swimmers, and not on viscosity, provided that the medium is a Newtonian fluid \cite{poon2013physics}. Thus, non-constancy of $\bar v/\bar \Omega$ indicates either changing cell geometry or non-Newtonian effects \cite{Martinez2014}. 

Note that the relative merits of DDM and real-space tracking for obtaining $\bar v$ and $\beta$ has been discussed in depth before \cite{MartinezDDM}. 

\subsection{Imaging protocol}

We use $d \approx \SI{400}{\micro\meter}$ deep flat glass sample cells filled with $\approx \SI{150}{\micro\litre}$ of suspension and rendered air tight with petroleum jelly to prevent drift due to evaporation and stop replenishment of \ce{O2} after it is exhausted by respiring bacteria.\footnote{See Section~\ref{sec:smallmol_t}, especially Fig.~\ref{fig:glucose}(b), to see why we want no \ce{O2} supply.} DDM shows that contact with petroleum jelly does not change $(\bar v,\beta)$. 

Movies in phase-contrast illumination (PF10$\times$/0.3 at 100 frames/s, $\sim 4000$~images at 512$\times$512 pixels $\Rightarrow \,\sim \SI{40}{\second}$ movies) are recorded in an inverted microscope (Nikon Ti) with a Mikrotron high-speed camera (MC 1362) and frame grabber (Inspecta 5, 1~Gb memory) at $22\pm 1\si{\celsius}$. Custom LabVIEW recording software controlling the microscope stage allows scanning of many samples (giving good averages) repeatedly (giving time resolution), giving data such as Figs.~\ref{fig:height}, \ref{fig:endo_speed} and \ref{fig:glucose}.

\subsection{Observational complications}

The majority of experiments to date using synthetic swimmers produced data at or near surfaces.  {\it Escherichia coli} cells can swim in the bulk for extended time, although cells encountering a wall tend to be trapped there \cite{TangSurface,Lauga2008}.   If performed sufficiently far from walls, DDM yields a genuine three dimensional speed distribution \cite{WilsonDDM,MartinezDDM} via fitting an ISF to Eq.~\ref{eq:DDM1}. Near a wall, the swimming is two-dimensional, and the $\sin(qv\tau)/qv\tau$ in Eq.~\ref{eq:DDM1} is replaced by $J_0(qv\tau)$, the zeroth-order Bessel function. However, it may not be possible to distinguish between these functional forms unless $P(v)$ is sufficiently narrow. Moreover, \ecoli\, swims in circles \cite{LaugaCircle}, giving $q$-dependent fitted dynamical quantities if the circle radii are comparable to the accessible range of $2\pi/q$. The complex motion of cells `tethered' to walls further complicate data analysis. We therefore typically image at \SI{100}{\micro\meter} away from the bottom of our capillaries (cf.~total cell length $L \lesssim \SI{10}{\micro\meter}$). 

\begin{figure}
\begin{center}
\includegraphics[width=0.4\textwidth]{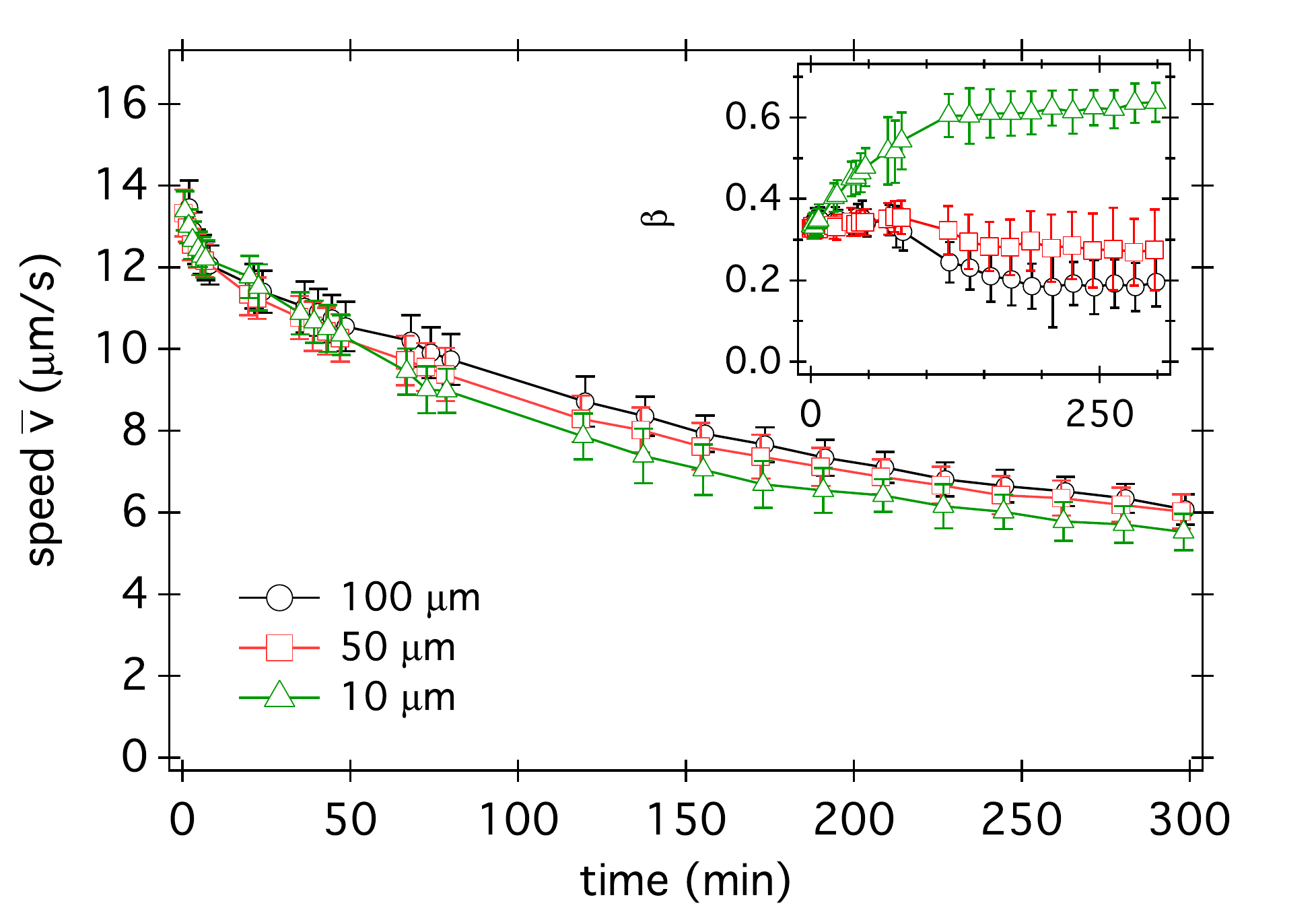}
\caption{Average swimming speed $\bar{v}$ (main) and non-motile fraction $\beta$ (inset) vs. time at three heights from bottom wall for WT AB1157 at $5\times 10^8$~cells/ml.}
\label{fig:height}
\end{center}
\end{figure}

The sedimentation height of non-motile cells, Eq.~\ref{eq:sedheight}, can be enhanced by 50\% or more because swimmers enhance their diffusivity \cite{Jepson2013}. Nevertheless, $l_g \approx \mathbf{O}(\SI{10}{\micro\meter})$, so that non-swimmers will accumulate on the bottom on a time scale of $d/v_s \lesssim \SI{70}{\minute}$ (using $v_s \approx \SI{0.1}{\micro\meter\per\second}$, Eq.~\ref{eq:sedspeed}). Measurement of WT swimmers \SI{10}{\micro\meter} from the bottom capillary surface, Fig~\ref{fig:height} inset, shows that $\beta$ increases with time, until it saturates just over an hour ($\approx d/v_s$) into the measurement. All non-motile cells have now sedimented into a layer of thickness $\approx l_g$. Correspondingly, $\beta$ is constant at any single bulk position (e.g., 50 or \SI{100}{\micro\meter} from the bottom, Fig.~\ref{fig:height} inset) until the sedimentation front passes (at some fraction of $d/v_s$), whereupon $\beta$ drops.  

Measurement of WT swimming \ecoli\, in a sealed capillary at our standard height of \SI{100}{\micro\meter} from the bottom shows that $\bar v$ decrease with time, Fig.~\ref{fig:height}. A priori, it is conceivable that this is because fast swimmers encounter surfaces and become trapped \cite{TangSurface,Lauga2008} more rapidly, thus leaving slower swimmers in the bulk. However, measurements at three heights, Fig.~\ref{fig:height}, give essentially identical results,\footnote{Note the longer time axis here than those shown later in Figs.~\ref{fig:endo_speed} and \ref{fig:glucose}.} so that the observed decreasing $\bar v(t)$ is {\it not} due to gradual kinematic accumulation at surfaces. 

Drift gives ballistic motions that masquerade as swimming. Evaporative drift can be eliminated by proper sealing, but unavoidable drift occurs at short times due to loading. These transients decay on a time scale $\gtrsim \rho L^2/\eta$ for a liquid with viscosity $\eta$ and density $\rho$, where $L$ is a characteristic length \cite{Guyon}. We used a $\SI{5}{\centi\meter} \times \SI{8}{\milli\meter} \times \SI{0.4}{\milli\meter}$ capillary throughout. The \SI{8}{\milli\meter} dimension leads to resolvable speed changes in our time window on the  $\approx \SI{1}{\minute}$ time scale, which probably accounts for the initial fast decay in $\bar v(t)$, Fig.~\ref{fig:height}. We find that decreasing this dimension gives faster initial decay. 

\subsection{Fitting lore}

The decay of the ISF, Eq.~\ref{eq:DDM2}, is due to diffusion (parameterised by $D$) and ballistic motion (parameterised by~$v$). Satisfactory fitting of the data in the form of $g(q,\tau)$ using Eq.~\ref{eq:DDM2} and Eq.~\ref{eq:DDM1} to obtain $D$ and $v$ depends on convincingly decoupling these contributions, which is straightforward when the two decays have well-separated characteristic times, Fig.~\ref{fig:easy_difficult_fitting}. Overlap of these two contributions occurs when their relaxation times approach, i.e. $v \rightarrow qD$, which, for WT $E.~ coli$ in bulk, $D\gtrsim 0.3~\mu$m/s, and $q \lesssim 2.2~\mu\text{m}^{-1}$ (using a $10\times$ objective), occurs at $v\lesssim 1~\mu$m/s. However, due to the distribution of speeds, this limit is underestimated. In practice, we can resolve $\bar v$ down to 2 to 3 \si{\micro\meter\per\second} in BMB, Fig.~\ref{fig:easy_difficult_fitting}. Note also that if $\beta\rightarrow 0$ or 1, the fitted value of $D$ or $v$ respectively will be subject to large errors. 

\subsection{Motility plates}

\begin{figure}
\begin{center}
\includegraphics[width=0.38\textwidth]{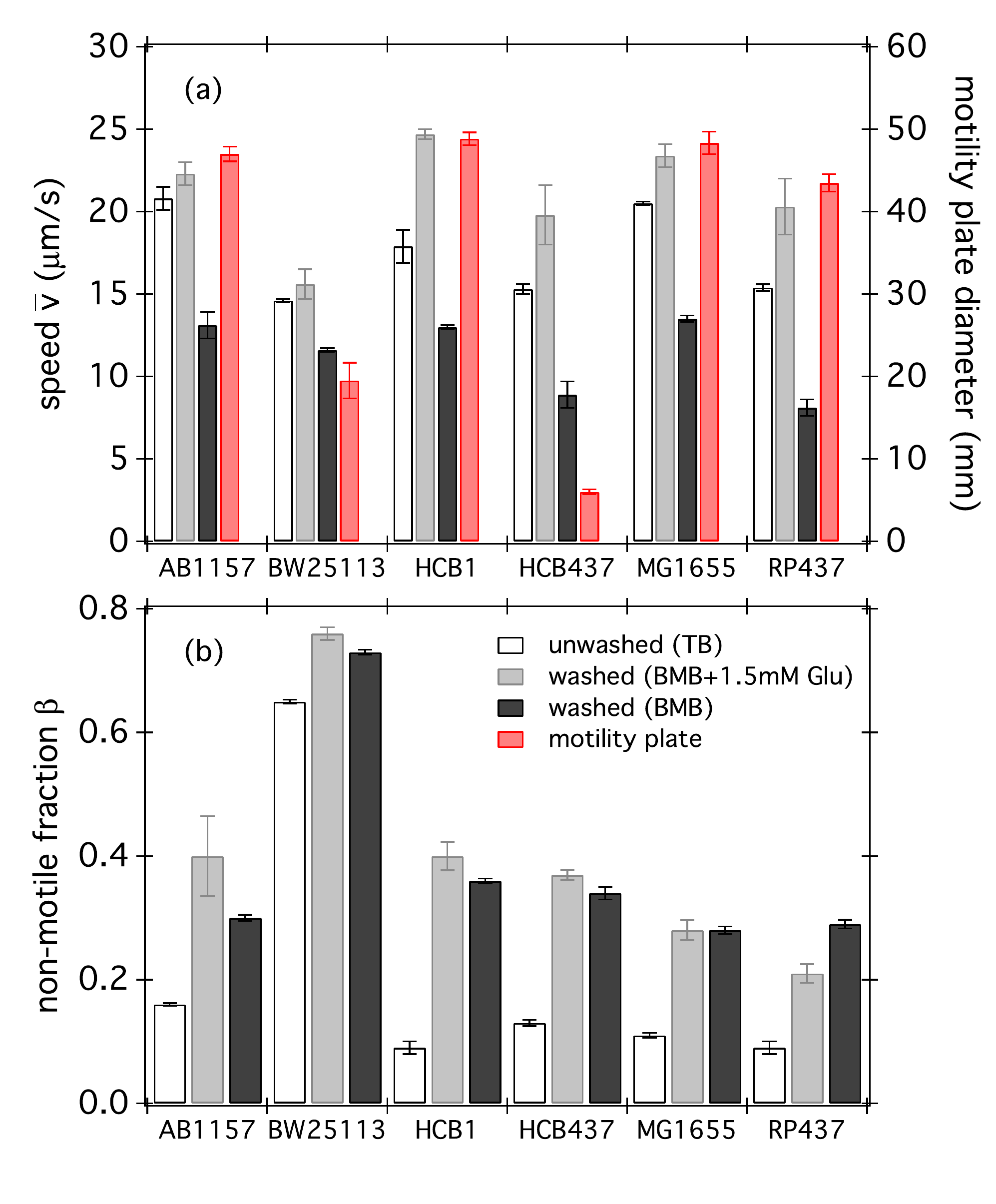}
\caption{(a) Average swimming speed $\bar{v}$ (left axis) and (b) non-motile fraction $\beta$ for selected \ecoli\, coli strains, from one batch culture in each case, before washing (in TB, white) and after washing (in BMB+1.5~mM glucose, grey or in BMB, black) at a cell density of $5\times 10^8$cells/ml. Red bars in (a) are diameters of solid, circular colonies in 22-hour LB motility plates (right axis). Note that for HCB437, the cells did not penetrate the agar but stayed on the surface.}
\label{fig:histogram}
\end{center}
\end{figure}

When chemotactic strains are inoculated into `soft agar' ($\gtrsim~0.2$~wt\%), they spread out in rings \cite{Wolfe1989}. Non-chemotactic mutants \cite{Wolfe1989} and  the chemotactic WT when the agar is $\gtrsim 0.3$wt\% \cite{Croze}  spread out instead as solid discs. The diameters of these rings or solid discs are often used to assay motility.

We inoculated \SI{5}{\micro\litre} drops of OD = 0.3 TB-grown \ecoli\, into 0.3wt\% LB agar plates, measured the diameter of the solid colony discs after \SI{22}{\hour} of incubation at \SI{30}{\celsius}, and compared these with $\bar v$ measured from DDM in three different media, Fig.~\ref{fig:histogram}(a). For chemotactic WT cells, there is reasonably correlation between motility plates and DDM measurements in all three media, but especially glucose-supplemented BMB.  

The non-chemotactic spreading of growing cells can be modelled using the Fisher equation \cite{Murray2002}. This predicts a front speed of $u \sim 2\sqrt{D\alpha}$, where $\alpha$ is the growth rate, and $D$ is the cells' effective diffusivity. The latter is  $\approx \bar v^2 \tau/3$, with $\tau$ being the average time between tumbles \cite{Loveley}. The colony diameter after time $\Delta t$ is $d = u\Delta t \sim 2\bar v \Delta t \sqrt{\alpha\tau/3} \propto \bar v$. Using the value of $\tau \sim \SI{1}{\second}$ in bulk media \cite{BergBrown} and the glucose-BMB data for HCB1, Fig.~\ref{fig:histogram}(a), we find $\alpha \approx \SI{0.0003}{\per\second}$, or a doubling time of $\alpha^{-1}\ln 2 \gtrsim \SI{30}{\minute}$, which is reasonable under our conditions. 

Two caveats are in order. First, if the agar concentration allows chemotaxis, the dependence of the speed of the front (now a `ring') on $\bar v$ becomes more complex \cite{Croze}. Secondly, we find that HCB437, a smooth swimmer, did not penetrate the agar; its minimal spread on the plate surface does {\it not} reflect $\bar v$ in bulk media. Thus, motility plates must only be used with caution.

\section{Preparing motile \textbf{\textit{E. coli}}}
\label{sec:culture}

We turn now to culturing bacterial swimmers. Readers needing to learn generic microbiological protocols should use standard manuals \cite{bench,pocket} and consult a microbiologist. We focus on extra procedures for obtaining {\it motile} cells. 

\subsection{Culturing cells}

Our protocol is based on that of H.~C.~Berg (Harvard), and involves Luria broth (LB, originally `lysogeny broth' \cite{Bertani2004}) and tryptone broth (TB). TB is a mixture of amino acids from hydrolysing milk casein proteins; LB is tryptone broth plus yeast extract, the latter containing various carbohydrates not in TB. 

We grow single colonies from frozen stocks on LB agar plates at \SI{30}{\celsius} overnight.  A single colony is transferred from a plate to \SI{10}{\milli\litre} of liquid LB and incubated overnight ($\approx \SI{16}{\hour}$) at \SI{30}{\celsius}\footnote{Growth is optimal at $T=\SI{37}{\celsius}$, but lower $T$ promotes swimming \cite{Berg,AdlerEnviron}.}, shaken (for aeration) at 200~rpm. Finally, cells are diluted 100-fold into \SI{35}{\milli\litre} of TB and grown for $\sim \SI{4}{\hour}$ at \SI{30}{\celsius} shaken at 200rpm. DDM shows that transferring to TB from LB results in higher $\bar v$. As cell sizes (Table~\ref{tab:length}) and other properties change with growth phase, it is important that growth times for different  experiments should be kept constant. 

\subsection{Washing cells}
\label{sec:wash}

The next step is to transfer cells from TB to a `minimal medium' with no exogenous nutrients. This prevents growth, which is an unwanted complication in the long run. More importantly, we find that cells use oxygen very quickly in TB, and once oxygen is exhausted, $\bar v$ decreases dramatically. For WT AB1157 at OD = 0.3, this occurs within \SI{10}{\minute}, which is not long enough for experiments. 
We therefore wash (by filtration) and transfer cells into Berg's motility buffer (BMB), containing 6.2~mM \ce{K2HPO4}, 3.8 mM \ce{KH2PO4}, 67 mM \ce{NaCl}, and 0.1mM EDTA. Cells undergo 4 successive filtrations (= 3 washes, see \ref{app:filter}) until the concentration of TB is diluted $10^3$-fold by BMB. After the last wash, the suspension on top of the filter is removed and transferred to a 50ml centrifuge tube together with the filter, where cells are carefully suspended by rolling the tube. This  yields 1-3~\si{\milli\litre} of highly concentrated cells (OD~$\sim10$), allowing the preparation of many dilute samples. 

\begin{figure}
\begin{center}
\includegraphics[width=0.35\textwidth]{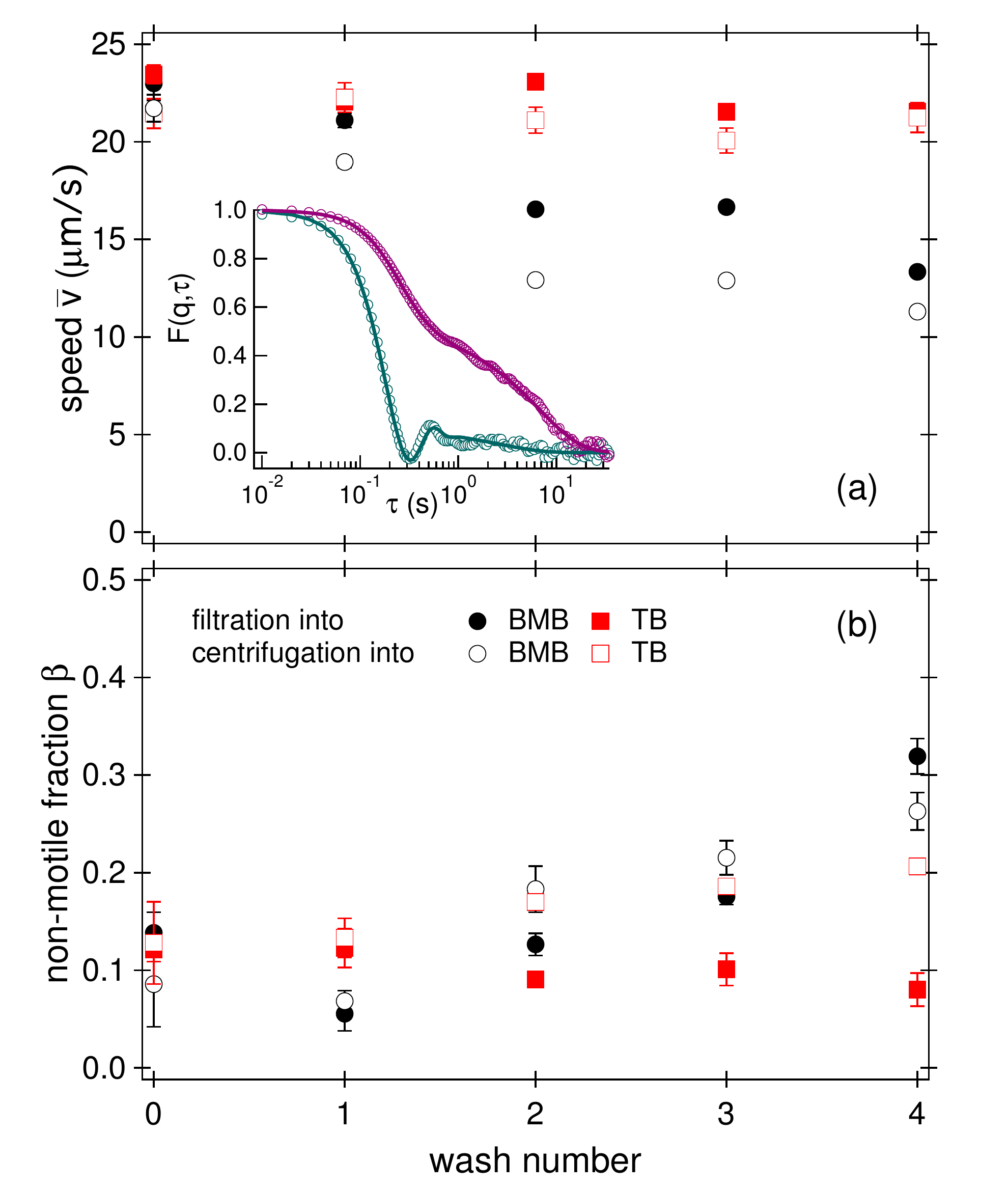}
\caption{(a) Average swimming speed $\bar{v}$ and (b) non-motile fraction $\beta$ during a washing process from TB to BMB (black) and from TB to TB (red) of WT AB1157, using filtration (filled) or centrifugation (open) at final cell density of $5\times 10^8$cells/ml. Inset: measured (symbols) and fitted (line) $f(q,\tau)$ of these cells in TB before washing (turquoise, lower) and in BMB after 4 washing steps (purple, upper). Data sets are from one batch culture.}
\label{fig:wash}
\end{center}
\end{figure}

Transfer from TB into BMB decreases $\bar v$ and increases $\beta$ for all strains studied, Fig.~\ref{fig:histogram}. This change in cellular motility is reflected in the ISF measured by DDM, Fig.~\ref{fig:wash}(a) inset. In TB, the population of mostly motile cells (a single decay) has a narrow speed distribution (oscillations). In BMB, two processes are clearly visible: a sizeable fraction of non-motile (diffusive) cells coexists with (ballistic) swimmers, whose speed distribution is now considerably wider (no oscillations). 

Figure~\ref{fig:wash} also shows that washing by filtration has non-trivial effects on $\bar v$ and $\beta$, probably due to a combination of mechanical damage to flagella, regrowth (possible in TB) and change of medium. Pipetting may also have an effect. After 20 pipetting cycles in BMB using a \SI{0.6}{\milli\meter} tip at normal expulsion speeds, we find that $\bar v$ drops by $\approx 50\%$ and $\beta$ increases by a factor of $\lesssim~3$. It is not until the tip diameter has been increased to $\approx \SI{2}{\milli\meter}$ that the effect on $(\bar v, \beta)$ begins to saturate. Such flow-driven damage of \ecoli\, flagella has been noted before \cite{Turner2012}. 

If low cell concentrations are required or many strains need to be processed simultaneously, we use a bench-top centrifuge to wash small volumes ($\lesssim \SI{2}{\milli\litre}$) 3 times at moderate speeds (\SI{2}{\minute} at 8000~rpm) to produce $\approx \SI{1}{\milli\litre}$ of OD $\approx 0.3$ cell suspension. Figure~\ref{fig:wash} shows that this procedure produces populations with higher $\beta$ in TB, but it is considerably faster than filtration. 

\subsection{Counting cells}
\label{sec:count}

\begin{figure}[t]
\begin{center}
\includegraphics[width=0.35\textwidth]{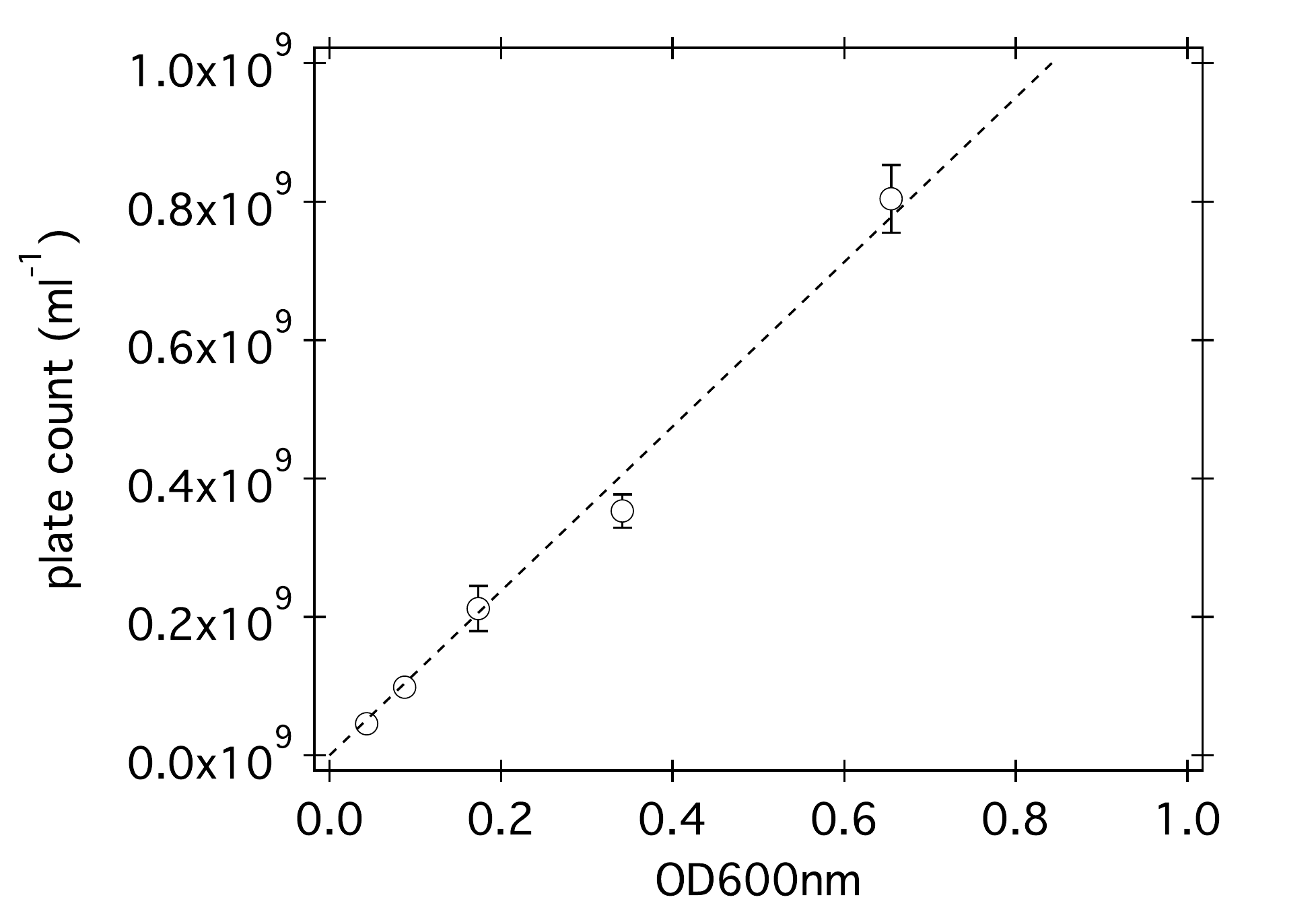}
\caption{Example of viable plate count vs.~OD used for calibrating the latter. These values are typical for our culture protocol.}
\label{fig:plate}
\end{center}
\end{figure}

The most common protocol for relating OD to cell number is the `viable plate count' \cite{bench,pocket}. Briefly, a sample is diluted to the relevant OD range (say, 0.05-1) using phosphate buffered saline (PBS).\footnote{PBS = 137~mM \ce{NaCl}, 2.~mM \ce{KCl}, 10~mM \ce{Na2HPO4} and 2~mM \ce{KH2PO4}.} One part of the sample is used to measure OD; another part is used to prepare tenfold serial dilutions (down to $10^{-6}$) by transferring \SI{100}{\micro\litre} sample to \SI{900}{\micro\litre} PBS with vigorous mixing between each dilution. \SI{100}{\micro\litre} of the final two dilutions are spread on replicate LB plates and incubated at \SI{37}{\celsius} overnight. Next day, individual colonies can be seen. Each of these has arisen from a `colony forming unit' (CFU), a shorthand for `probably but not certainly a single cell'. These are counted to arrive at the viable plate count in units of \si{\mbox{CFU}\per\milli\litre}. The cell density of the original sample is then calculated taking into account the dilution. Plotting the OD vs the cell density from plate counting, Fig.~\ref{fig:plate}, gives the desired calibration.

\section{\textbf{\textit{E. coli}} energetics}
\label{sec:energetics}

Self-propulsion requires energy. The energetics of synthetic swimmers is partially understood \cite{Seifert2012,Velegol2013}. Here we give a  simplified introduction to \ecoli\, bioenergetics. Details can be found in textbooks \cite{NeidhardtText,Stryer,Moat2002,McMurray2005,Kim2008,Nicholls2013}.\footnote{An unusual, `coarse-grained' introduction is given by Nobel laureate Christian de Duve \cite{deDuve}. See also  tutorials at \cite{ColiStudent} under `The Microbe' tab.}

Bioenergy comes from transferring electrons from high-energy to low-energy bonds through a sequence of molecules. The energy is used to pump protons out of the cell, giving rise to a proton motive force (PMF), which powers swimming. In aerobic respiration, the electron donors are various reduced foods rich in \ce{C\bond{-}H} bonds, while the final electron acceptor is typically oxygen. The electron pair in the \ce{C\bond{-}H} bond lowers its energy as it is transferred (notionally as \ce{:H-}) to an \ce{O\bond{-}H} bond (in water), where, because \ce{O} is more electronegative than \ce{C}, the electron pair interacts more strongly with positively charged nuclei than in \ce{C\bond{-}H}, and therefore has lower energy. The cell harnesses the energy released for chemical work. 
The `favourite' supplier of high energy electrons for \ecoli\, is $\alpha$-D-glucose (\ce{C6H12O6}).\footnote{Glucose polymers constitute $> 50\%$ of dry terrestrial biomass \cite{McMurray2005}.} The full oxidation of this molecule in aerobic respiration generates $\approx \SI{2.8}{\mega\joule\per\mol}$ under physiological conditions 
\begin{equation}
\ce{{\scriptsize \setcrambond{2pt}{}{}\chemfig{HO-[2,0.5,2]?<[7,0.7](-[2,0.5]OH)-[,,,,line width=2pt](-[6,0.5]OH)>[1,0.7](-[6,0.5]OH)-[3,0.7]O-[4]?(-[2,0.3]-[3,0.5]OH)}}
 + 6O2 -> 6CO2 + 6H2O}\,\,\Delta{G} \approx -10^3k_BT, \label{eq:stoichem}
\end{equation}

\begin{figure}
\begin{center}
\includegraphics[width=0.475\textwidth]{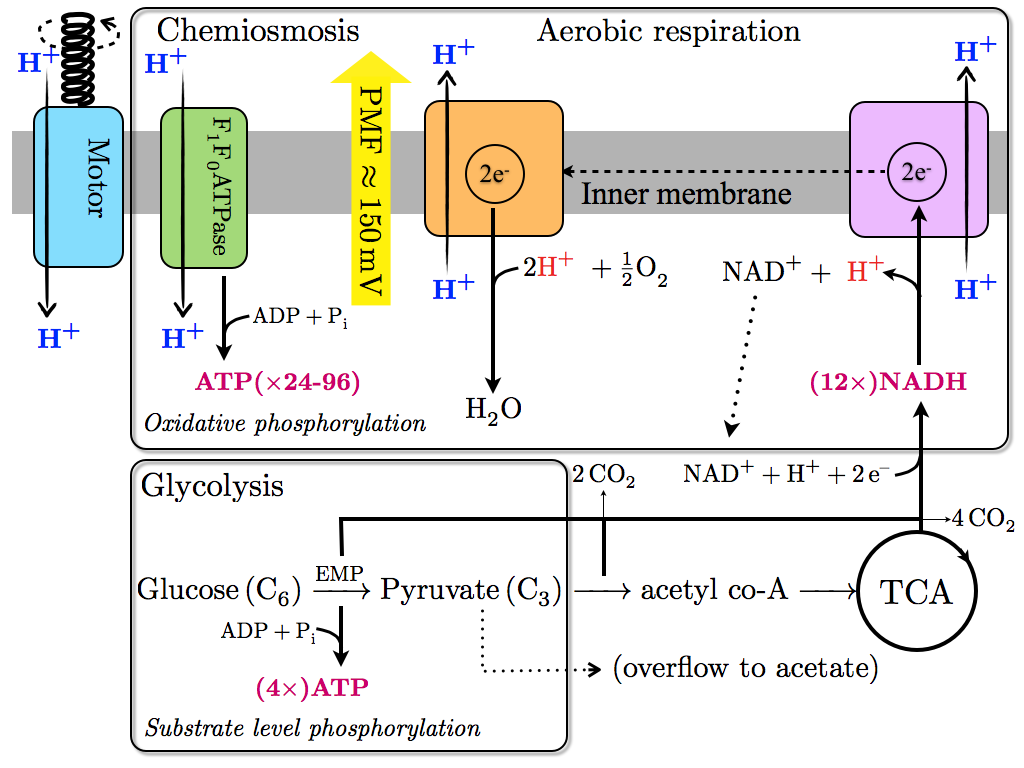}
\caption{Schematic of \ecoli\, bioenergetics, with high-energy compounds ATP and NADH picked out in bold. Glycolysis breaks down glucose (\ce{C6}) into two pyruvates (\ce{C3}), producing \ce{4ATP} by substrate-level phosphorylation. Pyruvate is reduced to acetyl co-A with the emission of \ce{CO2}. (Pyruvate is catabolised to acetate when glucose is abundant.) Acetyl co-A enters the tricarboxylic acid (TCA) cycle to produce \ce{12NADH} per glucose. Each NADH donates it high-energy electron pair, \ce{2e-}, to respiratory enzymes in the inner membrane, which pass them to a terminal acceptor, here \ce{O2}, which is reduced to water, regenerating \ce{NAD+} (oxidised NADH) for glycolysis (dotted arrow).  The drop in energy as \ce{2e-} passes along the respiratory enzyme chain is used to pump \ce{H+} out of the cytoplasm, generating a proton motive force (PMF), which drives protons through rotary flagella motors and through \ce{F1F0}ATPase to make ATP from ADP (chemiosmotic oxidative phosphorylation).  }
\label{fig:PMF}
\end{center}
\end{figure}

Figure~\ref{fig:PMF} summarises the aerobic catabolism (break down) of glucose, a \ce{C6} compound. First, glycolysis via the Embden-Meyerhof-Parnas (EMP) pathway\footnote{When glucose is the sole carbon and energy source for \ecoli, $\approx 75\%$ of it channels into the EMP pathway; the rest feeds into the hexose monophosphate (HMP) pathway, which mainly provides precursors for biosynthesis, i.e. anabolism \cite{Kim2008}. However, the HMP pathway can also feed pyruvate into the TCA cycle \cite{Kim2008}. A fraction of TCA intermediates supplies anabolic precursors. Such `anabolic syphons' together with overflow metabolism, reaction \ref{eq:overflow}, mean that the stoichiometry in reaction \ref{eq:stoichem} is never satisfied on average. \label{note:stoichem}}  breaks glucose into two C$_3$-halves, pyruvates (\ce{CH3COCOO-}), a key metabolic intermediate. Then, pyruvate is oxidised to acetyl coenzyme A (acetyl co-A), another key metabolic intermediate, and fed into the `tricarboxylic acid cycle' (TCA = Krebs or citric acid cycle) to be further broken down. After glycolysis, \ce{CO2} is produced without molecular \ce{O2} through the progressive enzymatic dehydrogenation of fragments of catabolised glucose. 

During glucose catabolism, the energy released is stored in high-energy compounds, mainly adenosine triphosphate (ATP) and (reduced) nicotinamide adenine dinucleotide (NADH). ATP is generated from adenosine diphosphate (ADP) by the addition of an inorganic phosphate group:
\begin{equation}
 \ce{ADP + P_i -> ATP}.
\end{equation}
NADH comes from reducing \ce{NAD+}, the oxidised form of this compound, partly via \ce{CO2}-producing dehydrogenations:
\begin{equation}
\schemestart \ce{NAD+  + H+ + 2e-} \arrow{-U>[ \ce{COOH}][\small \ce{CO2}]} \ce{NADH} \schemestop.
\end{equation}

The generation of 4ATP per glucose {\it en route} to pyruvate is known as `substrate level phosphorylation' (SLP), where phosphate groups are added to ADPs without using molecular \ce{O2}. The 12~NADH generated in glycolysis and the TCA cycle give rise to up to 96 ATPs through the process of `oxidative phosphorylation'. NADH donates its energetic electron pair to respiratory enzymes in the inner membrane, ultimately passing them to \ce{O2} as `terminal electron acceptor', producing \ce{H2O} and regenerating \ce{NAD+} for glycolysis and the TCA cycle. Without such regeneration, e.g., because respiratory enzymes are poisoned or because \ce{O2} runs out, the TCA cycle falters.

The energy released as \ce{2e-} drops down the potential energy ladder of respiratory enzymes pumps between 2 to 8 \ce{H+} out of the cell (8 at high [\ce{O2}]) \cite{Unden1999}, generating a proton motive force (PMF) of $\approx \SI{150}{\milli\volt}$. Driven back to the cellular interior by the PMF,  protons do work in active solute transport, turning flagella motor and making ATP in `rotary enzyme' F$_1$F$_0$-ATP synthase (`chemiosmosis', requiring $\approx \ce{4H+}$ per ATP).

When glucose is plentiful, it is incompletely oxidised to acetate (\ce{CH3COO-}) via pyruvate:
\begin{equation}
\ce{C6H12O6} + \ce{2O2} -> \ce{2CH3} \chemfig{C(-[7,0.7]\ce{O^{-}H^{+}})=[1,0.7]O} + \ce{2CO2} + \ce{2H2O}. \label{eq:overflow}
\end{equation}
Such `overflow metabolism' (or the `Crabtree effect') generates less energy, and excretes acetate for potential later reabsorption and further oxidation when glucose is less plentiful \cite{Vemuri2006,Noack2012}.

How \ecoli\, adapts to decreasing [\ce{O2}] in its environment is complex \cite{Thomas1972,Neidhardt1983a,Neidhardt1983b}. At [\ce{O2}]~=~0 and without other terminal electron acceptors, \ecoli\, ferments glucose into acetate, formate, lactate and succinate, using metabolic intermediates as electron acceptors and producing ATP by SLP, with acetate production generating most ATP~\cite{Thomas1972,San2005}.  In such `mixed-acid fermentation', respiratory enzymes are inactive, and other pathways maintain the PMF, including F$_1$F$_0$-ATPase working in reverse, hydrolysing ATP to ADP and using the energy to pump protons out of the cell \cite{Moat2002,Nicholls2013,Neidhardt1983a}. The PMF powers flagella motors. Within a wide range, the motor speed is proportional to the PMF \cite{Gabel2003}. If the PMF is short circuited (e.g. by making the membrane permeable to protons) \cite{Berg1977}, or drops to zero when `fuel' runs out, swimming ceases. 

Under starvation conditions (i.e., no exogenous nutrients), such as occurs in BMB, \ecoli\, obtains energy by metabolising intracellular resources. Such `endogenous metabolism' is complex~\cite{Ribbons1963,Dawes1965} and poorly understood. For starved cells grown in TB, catabolism of free amino acids and free or RNA-derived ribose (a sugar) feeds into the TCA cycle \cite{McMurray2005}, so that \ce{O2} or other terminal acceptors are needed. 

\section{\textbf{\textit{E. coli}} swimming powered by endogenous metabolism}
\label{sec:smallmol}

{\it Escherichia coli} and other bacteria \cite{Harwood1989,Ford1999,Winter2013} can swim powered entirely by endogenous metabolism. We now probe the time dependence of $\bar v$. All data are obtained from DDM with WT AB1157 in BMB prepared using our standard protocol. 

\begin{figure}[t]
\begin{center}
\includegraphics[width=0.5\textwidth]{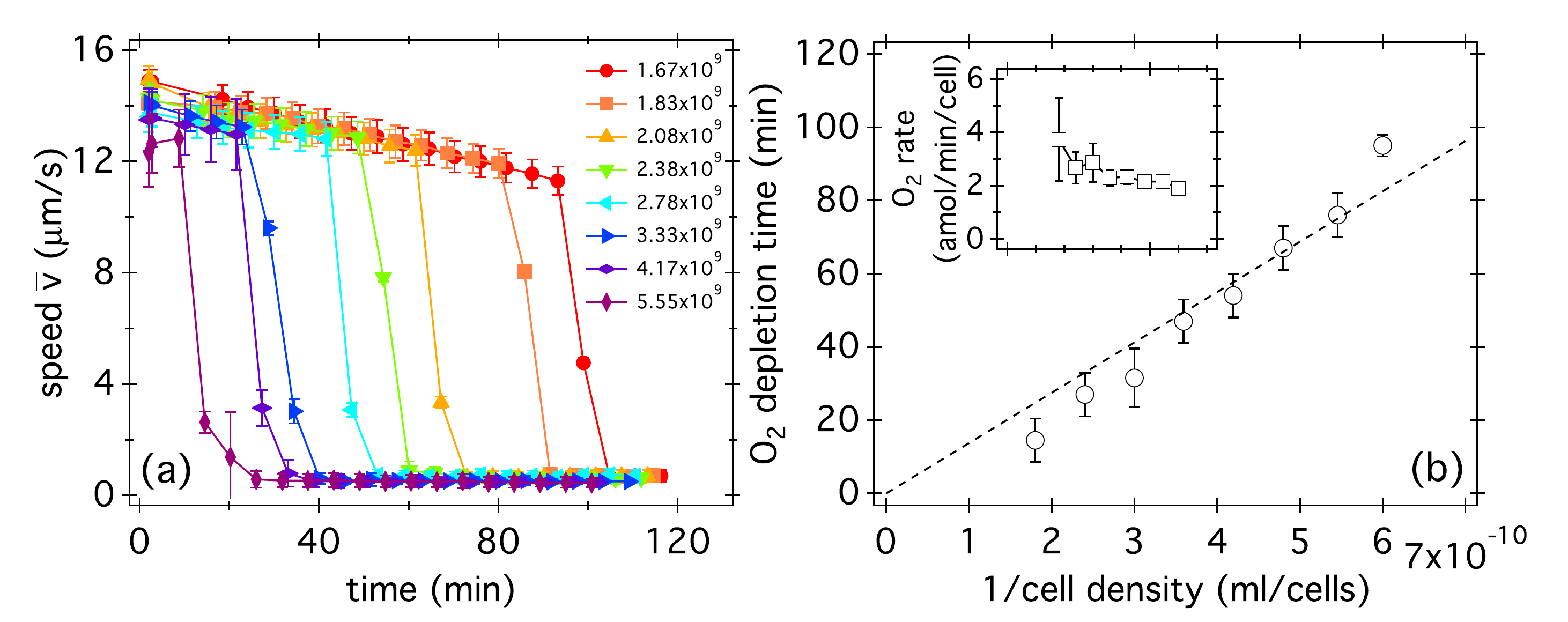}
\caption{(a) Average swimming speed $\bar{v}$ vs.~time for the capillary protocol in BMB at  cell densities indicated in the legend ($\si{\mbox{cells}\per\milli\litre}$). Lines guide the eye. The `crash' in $\bar{v}$ in each case is due to \ce{O2} exhaustion. Data sets from one batch culture. (b) Corresponding \ce{O2} exhaustion time as a function of inverse of cell density. Error bars relate to the accuracy of measuring the `crash' time.}
\label{fig:endo_speed}
\end{center}
\end{figure}

\subsection{Observations and practical implications} 
\label{sec:smallmol_obs}

Figure~\ref{fig:endo_speed}(a) shows the average swimming speed  as a function of time, $\bar{v}(t)$, in sealed capillaries at various cell densities, $n$. Cells slow down, approximately linearly with time. If $\approx 10\%$ slowing down is tolerable, then there is a useable window of $\approx$~1~h at $n = 2\times10^9\,\si{\mbox{cells}\per\milli\liter}$, which ends in a `crash'. 

The time of this `crash', $t_c$, drops with $n$; we suggest that it is due to \ce{O2} depletion. Quantitatively, $nt_c \approx$~constant, Fig.~\ref{fig:endo_speed}(b), as expected if the initial [\ce{O2}] and the \ce{O2} consumption rate per cell, $Q$, are invariant with  $n$. Assuming that the BMB was initially saturated with \ce{O2}, and using literature  solubilities~\cite{Carpenter1966}, which are only very weakly dependent on ionicity~\cite{Rasmussen2003}, we estimate $Q$ at each cell density, Fig.~\ref{fig:endo_speed}(b) inset, which is weakly dependent on $n$, rising from $Q_{\rm endo} \approx \SI{2}{\atto\mole\per\minute\per\mbox{cell}}$ at $n = 1.67 \times 10^9 \,\mbox{cells\,} \si{\per\milli\litre}$ to $Q_{\rm endo} \lesssim \SI{4}{\atto\mole\per\minute\per\mbox{cell}}$ at $n = 5.55 \times 10^9 \,\mbox{cells\,} \si{\per\milli\litre}$ (`a' $\equiv$ `atto' $=10^{-18}$). These values are comparable to those measured before using a different strain and under somewhat different conditions \cite{Ribbons1963}.\footnote{$Q_{\rm endo} \approx 10-20~\si{\milli\litre\per\hour\per\mbox{mg dry cell mass}}$~\cite{Ribbons1963} $\rightarrow$ 2-$\SI{4}{\atto\mole\per\minute\per\mbox{cell}}$ using \ce{O2} molar volume = \SI{24}{\deci\meter\cubed} and dry cell weight $\approx 0.3 \times 10^{-9}~\si{\milli\gram\per\mbox{cell}}$~\cite{NeidhardtText}.}

The highest density data in Figs.~\ref{fig:endo_speed} were obtained at $n \approx 5 \times 10^{9} ~\si{\mbox{cells}\per\milli\litre}$, or $\phi \approx 0.5\%$ of cell bodies. Extrapolating from $Q$ at this $n$, $t_c \approx \SI{1}{\minute}$ at $\phi \approx 5\%$; this is an upper bound, since $Q$ probably increases with $n$. Oxygen exhaustion therefore precludes experimentation at $\phi \gtrsim 1\%$ under these conditions. 

\begin{figure}
\begin{center}
\includegraphics[width=0.5\textwidth]{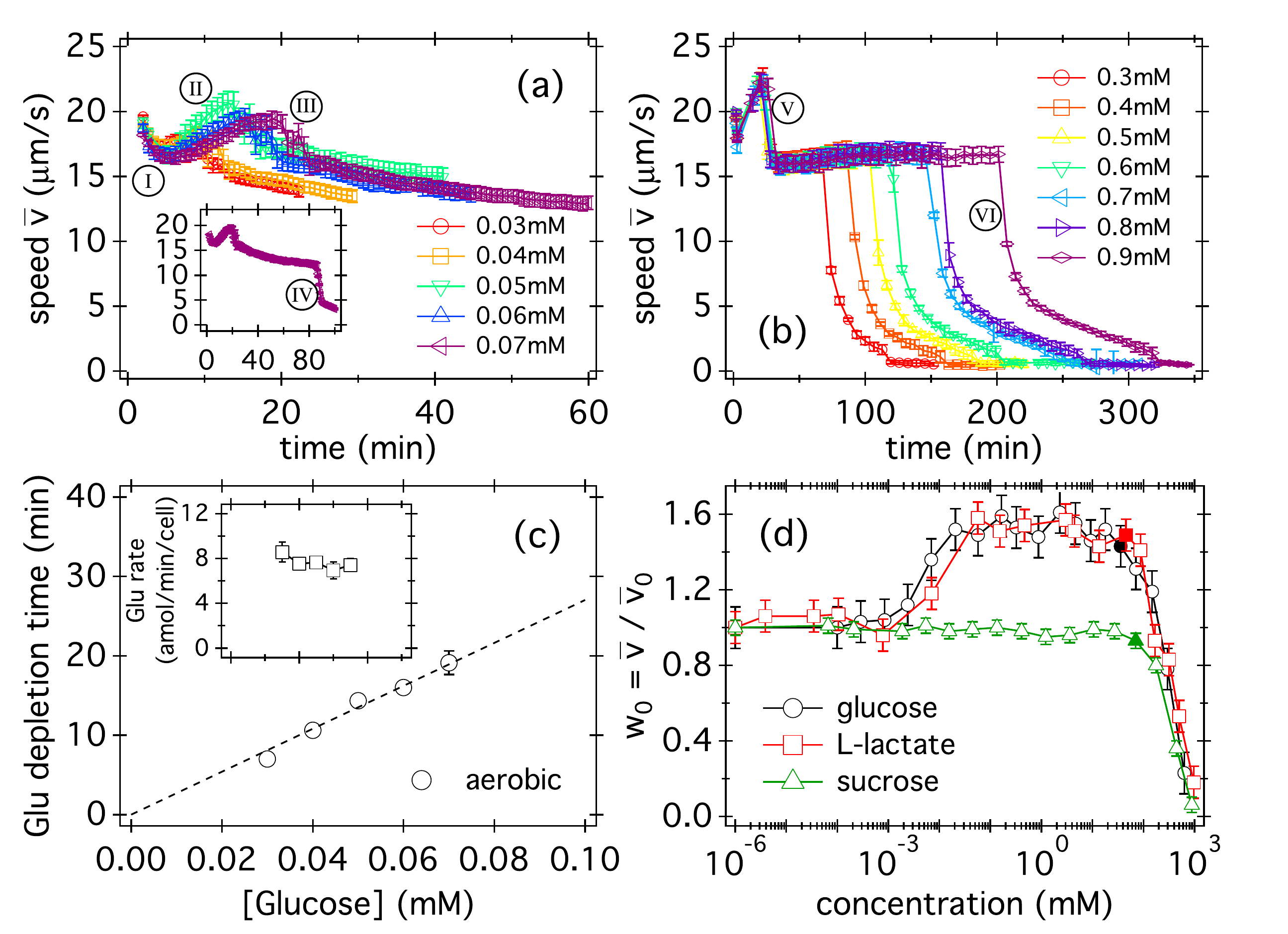}
\caption{Effect of glucose on time-dependent average swimming speed $\bar v(t)$ for WT AB1157 at $5\times~10^8$~cells/ml. (a) At low [Glu], glucose depletes first, causing the rapid drop  (III) in $\bar v$. Inset: a longer run at [Glu]=0.07~mM reveals a subsequent sharp drop (IV) due to \ce{O2} depletion. (b) At high [Glu], \ce{O2} depletes first, causing the first sharp drop (V) in $\bar v$. The second, larger sharp drop (VI) in $\bar v$ is due to glucose depletion. Datasets are from one batch culture. (c) Depletion time of glucose in aerobic condition versus glucose concentration.  Inset: aerobic glucose consumption rate versus [Glu]. (d) Speed $\bar{v}$ versus molecule concentration normalised to the speed in BMB, $\bar{v}_0$, for three small molecules, measured $\approx 5$~min after mixing and filling the capillary. Filled symbols are concentration markers for Fig.~\ref{fig:osmotic}. Error bars in (c) and (d) are standard deviations from averaging data from two and three independent batch cultures.}
\label{fig:glucose}
\end{center}
\end{figure}

\subsection{Physiological interpretation}

The observed slowing down in $\bar v$ at $t \lesssim \SI{1}{\hour}$ is probably not due to substrate exhaustion, as extrapolated literature data~\cite{Ribbons1963,Dawes1965} suggests this happens in $\approx \SI{10}{\hour}$. That $d\bar v/dt$ is identical at different $n$, Fig.~\ref{fig:endo_speed}(a), supports this suggestion. 

The ubiquitous signalling molecule cyclic di-guanylate (c-di-GMP) can slow down bacterial rotary motors by binding to the protein YcgR to form a molecular `brake'~\cite{Boehm2010}. However, we found that a $\mathit{\Delta}${\it YcgR} mutant behaved identically to the WT in sealed capillaries; so c-di-GMP is unlikely implicated. Instead, we infer a slow decrease in the PMF driving the rotary motors. The adaptive value and mechanism of this putative starvation response is unknown, but it can affect many PMF-dependent cellular functions, e.g. ATP synthesis by F$_1$F$_0$-ATPase. 

\section{\textbf{\textit{E. coli}} swimming powered by glucose}
\label{sec:smallmol_t}

If glucose is present, \ecoli\, utilises it first, and suppresses the expression of enzymes for processing other nutrients \cite{Gorke2008}. 

Glucose increases the swimming speed of \ecoli\, \cite{AdlerEnviron} (see also Fig.~\ref{fig:histogram}(a)) and other enteric bacteria \cite{Lai1997}. We now demonstrate how to use glucose to enable \ecoli\, AB1157 (cultured by standard protocol) to swim at a constant $\bar v$ for extended time. 

\begin{figure}[t]
\begin{center}
\includegraphics[width=0.35\textwidth]{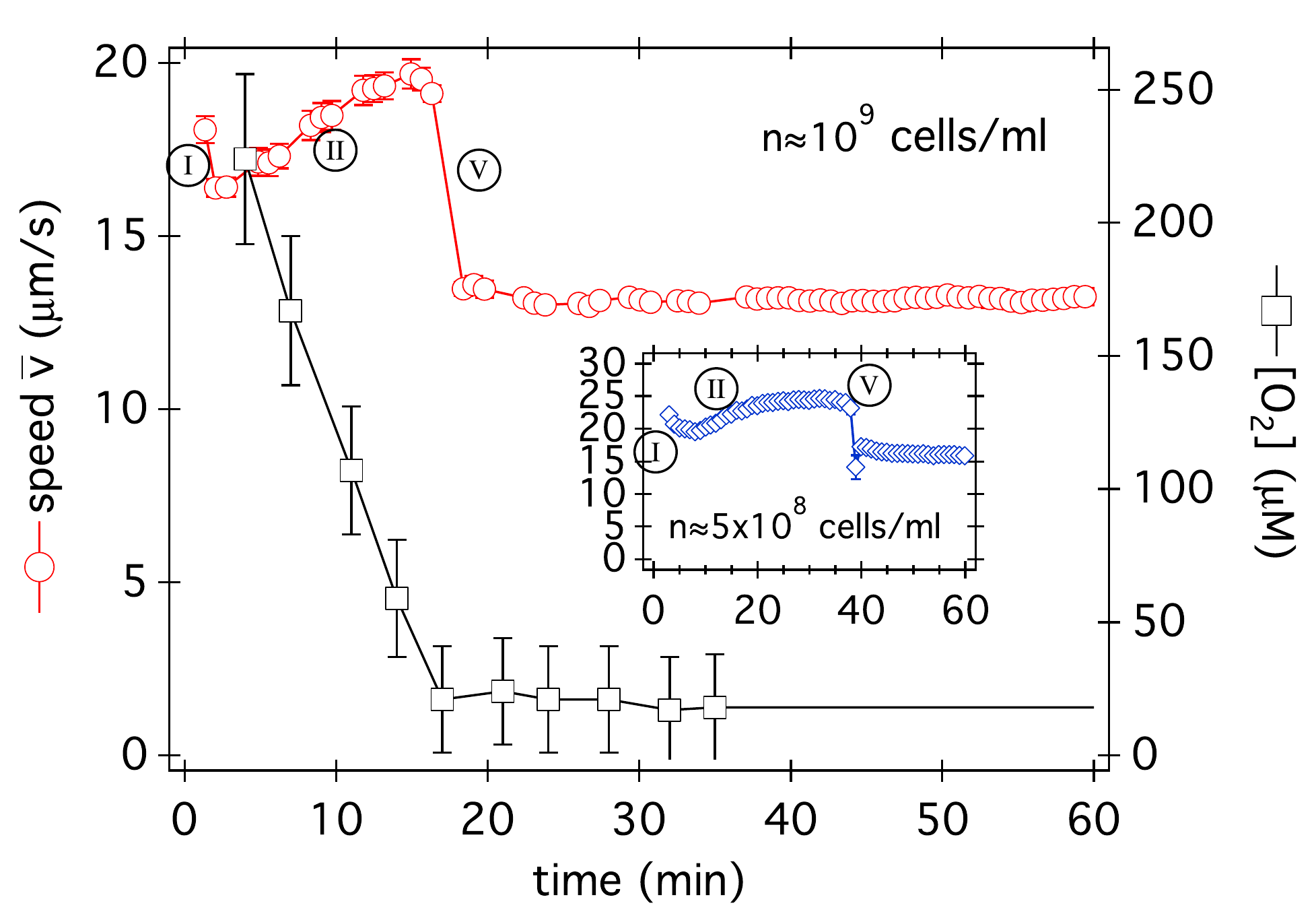}
\caption{Simultaneous measurements of average swimming speed $\bar{v}$ ({\color{red} $\circ$}, left axis) and [\ce{O2}] ($\square$, right axis), using fluorescence lifetime imaging microscopy (FLIM) vs. time in a sealed capillary for WT AB1157 at $\approx 10^9$~cells/ml and [\ce{Glu}]=0.5~mM. For FLIM, $\SI{47}{\milli\mbox{M}}$ of the \ce{O2}-sensitive dye ruthenium tris(2,2'- dipyridyl) dichloride hydrate (RTDP) was added to BMB (which did not affect the cells' motility). RTDP was excited using sub-ps pulses ($\lambda=450$nm, $\sim$1mW at 1MHz repetition rate) and its fluorescence imaged using a gated intensified CCD camera (Picostar HR-12QE, LaVision GmbH, Germany). The data was fitted with a single exponential decay yielding a (homogeneous) fluorescence lifetime map, which was averaged to estimate [\ce{O2}]. Inset: $\bar{v}$ at $5\times 10^8$~cells/ml to better show the saturation before \ce{O2} depletion. Data sets are from independent batch cultures with $\bar{v}(t=0)\approx \SI{12}{\micro\meter\per\second}$ and $\approx \SI{15}{\micro\meter\per\second}$, respectively. }
\label{fig:oxygen}
\end{center}
\end{figure}

\subsection{Observations and practical implications} 

Figure~\ref{fig:glucose}(a) and (b) show $\bar v$ at $n = 5\times 10^8 \, \si{\mbox{cells}\per\milli\litre}$ of WT AB1157 sealed in a capillary in BMB supplemented with low and high glucose concentrations respectively.  At $\SI{0.03}{\milli\mbox{M}} \leq [\mbox{Glu}] \leq \SI{0.07}{\milli\mbox{M}}$, Fig.~\ref{fig:glucose}(a), $\bar v$ rises gradually with time (II) after initial transients (I), until it drops suddenly (III). Thereafter, it decreases slowly with time. A longer run at $[\mbox{Glu}] = \SI{0.07}{\milli\mbox{M}}$,  Fig.~\ref{fig:glucose}(a) inset, reveals a second rapid drop (IV) that is sharper and of larger amplitude. The long, slow decrease with time and the final sudden drop are reminiscent of `endogenous propulsion', Fig.~\ref{fig:endo_speed}. Thus, the data suggest that glucose is consumed aerobically until it is depleted, whereupon cells revert to endogenous metabolism until \ce{O2} is, in turn, depleted. 

Consistent with this, the time of the first sharp drop (III) is proportional to [Glu], Fig.~\ref{fig:glucose}(c), and the deduced specific glucose consumption rate, inset, is nearly independent of [Glu], averaging to $\dot G\approx 6.9 \pm 0.5\,\si{\atto\mol\per\minute\per\mbox{cell}}$. We also measured $\dot G$ over a range of [Glu] biochemically (\ref{app:GluAssay}), and found a comparable value of $\dot G\approx 5 \pm 0.6\, \si{\atto\mol\per\minute\per\mbox{cell}}$.

Figure~\ref{fig:glucose}(c) implies that there will be a [Glu] at which \ce{O2} depletion will occur before glucose depletion. This indeed happens at the higher [Glu] explored in Fig.~\ref{fig:glucose}(b). Now, the very sharp, first `crash' (V) in $\bar{v}$ is due to \ce{O2} depletion, at which point there is still plentiful glucose in the medium. Consistent with this interpretation, the time of the first abrupt speed decrease (V) is independent of [Glu]. Again using literature \ce{O2} solubilities  \cite{Carpenter1966,Rasmussen2003}, we find an average \ce{O2} utilisation rate of $Q_{\rm Glu} = 22.6 \pm 0.6\,\si{\atto\mol\per\minute\per\mbox{cell}}$, which, as expected, is considerable higher than the $Q_{\rm endo} \approx$~2-$\SI{4}{\atto\mole\per\minute\per\mbox{cell}}$ reported by us (Section~\ref{sec:smallmol_obs}) and others~\cite{Ribbons1963} for endogenous metabolism.

Independent evidence for \ce{O2} depletion comes from monitoring the fluorescence life time $\tau$ of an \ce{O2}-sensitive dye \cite{Levine1997}. By measuring  $\tau$ and using our experimentally derived calibration $[\ce{O2}]=5\SI{68.1}{\micro\mbox{M}}\, (604$ns$/\tau-1)$ (consistent with \cite{Morris2007}), we can assay [\ce{O2}] in our sealed samples, Fig.~\ref{fig:oxygen}, which drops to $\approx 0$ precisely at the point of the abrupt decrease (V) in $\bar v$.

Subsequent swimming powered by anaerobic glucose metabolism shows $\bar v$ constant to $\pm \SI{1}{\micro\meter\per\second}$ until glucose depletion and $\bar v$ abruptly drops by a factor of 3 or more, Fig.~\ref{fig:glucose}(c) (VI). Thus, {\it cells in initially aerated BMB supplemented with enough glucose will swim at approximately constant average speed after \ce{O2} depletion and before glucose depletion.}

\begin{figure}
\begin{center}
\includegraphics[width=0.35\textwidth]{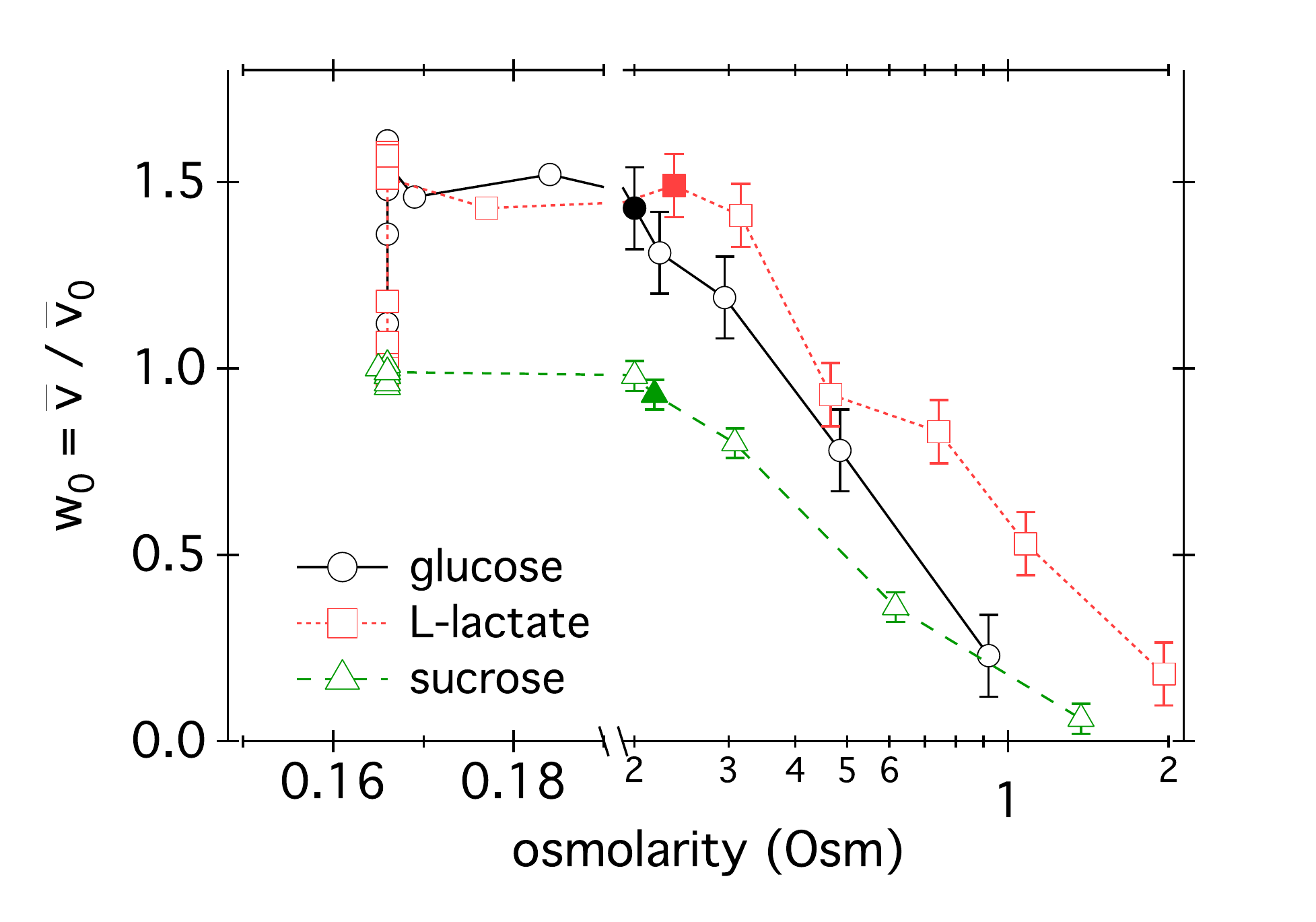}
\caption{Normalised average swimming speed, $\bar{v}/\bar{v}_0$, from Fig~\ref{fig:glucose}(b) as a function of the osmolarity. Note the bottom split axis with a linear scale from 150 to 190  mOsm (bottom left) and a log-scale from 200 to 2000 mOsm (bottom right). Filled symbols corresponds to markers in Fig.~\ref{fig:glucose}(d).
}
\label{fig:osmotic}
\end{center}
\end{figure}

The period of constant-$\bar v$ swimming can, at first sight, be arbitrarily extended by increasing [Glu], Fig.~\ref{fig:glucose}(b), up to saturation  ($\approx \SI{5}{\mbox{M}}$ at \SI{25}{\celsius}). However, a practical limit exists. Figure~\ref{fig:glucose}(d) plots $w_0$, the average swimming speed at various [Glu] normalised to the average speed at [Glu] = 0, both measured immediately after loading into sealed capillaries. At [Glu]~$\lesssim 10^{-3}\,\si{\milli\mbox{M}}$, glucose depletion occurs before measurements could begin, so that $w_0 = 1$. At [Glu]~$\gtrsim \SI{50}{\milli\mbox{M}}$, $w_0$ catastrophically drops. Repeating using L-lactate, which \ecoli\, can metabolise, and sucrose, which it cannot, also shows the same abrupt drop, which we believe is due to osmotic shock.

Figure~\ref{fig:osmotic} replots our data against solution osmolarity (determined using an Osmomat30, Genotec, Germany), showing that $\bar v$ always starts to decrease at an osmolarity between 0.2 and 0.4~Osmom, comparable to the osmolarities needed to start decreasing cell volumes in previous studies \cite{Pilizota2013}. Thus, there is a limit to the highest useable [Glu] for maintaining constant $\bar v$.\footnote{As expected from previous work \cite{Pilizota2013}, the results in Fig.~\ref{fig:osmotic} were time dependent: waiting longer produced partial recovery in the `crashed' speed. Nevertheless, a definite upper limit exists at $\approx 100$~mM small-molecule solutes.}

\subsection{Physiological interpretation}
\label{sec:glucose_physio}

Reaction \ref{eq:stoichem} for complete glucose oxidation requires an oxygen to glucose stoichiometry of $Q_{\rm Glu}/\dot G =6$. We measure $Q_{\rm Glu}/\dot G \approx 3.3 \pm 0.33$, i.e. about 50\% of the glucose consumed is fully oxidised. A combination of overflow metabolism and `anabolic siphon' (footnote \ref{note:stoichem}) probably explains this finding. 

Figure~\ref{fig:oxygen} shows the early-time portion of $\bar v(t)$ on an expanded time axis. After initial transients (I), $\bar v(t)$ rises (II) and then saturates (this is particularly clear in the inset) before the crash (V) due to \ce{O2} exhaustion. Immediately before these experiments, our cells have been harvested from TB, which contains little carbohydrates. While the operon of genes necessary for glucose transport ({\it pts}) is constitutively expressed irrespective of whether glucose is present, exposure to glucose can increase the expression of some {\it pts} genes by up to threefold \cite{Reuse1988,Postma1996}, which probably causes the rise (II) of $\bar v(t)$. 

Averaging over data sets, the ratio of the average speeds before and after the \ce{O2} `crash' (V) in the range 1.3 to 1.6. The PMF of K-12 \ecoli\, (strain AN387) grown aerobically on glycerol and fermentatively on glucose was previously measured to be \SI{-160}{\milli\volt} and \SI{-117}{\milli\volt} respectively  \cite{Unden1998}; $160/117 = 1.36$ is remarkably close to our speed ratio. 

\begin{figure}[t]
\begin{center}
\includegraphics[width=0.35\textwidth]{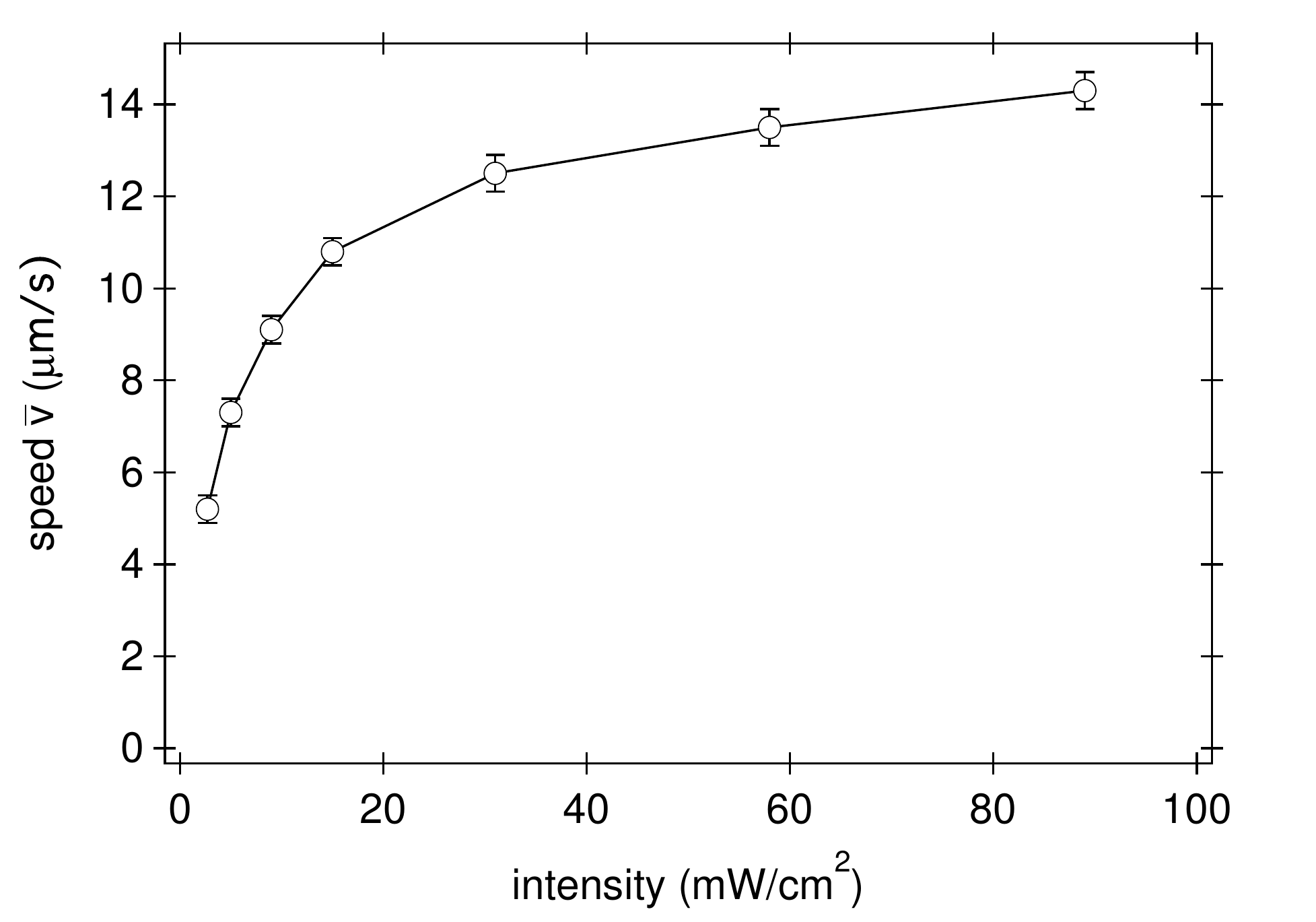}
\caption{Average steady-state swimming speed $\bar{v}$ of \ecoli\ MG1655 ($n = 10^9\,\si{\mbox{cells}\per\milli\litre}$) expressing SAR86 $\gamma$-proteorhodopsin at different intensities of incident green light ($\lambda = 510-\SI{540}{\nano\meter}$) after \ce{O2} depletion; data obtained starting from the highest intensity. }
\label{fig:LightControl}
\end{center}
\end{figure}

\subsection{How much energy does swimming use?}

The detailed mechanics of a peritrichous swimmer (= many flagella distributed over the cell body) is not yet understood \cite{BergTorque}, so that the vast majority of the literature on \ecoli\, is couched in terms of propulsion by a single `effective flagellum'. The power developed by the `effective motor' turning this `effective flagellum' has been estimated to be $P \approx 4 \times 10^{-16}\, \si{\watt}$ for swimming in minimal medium \cite{WuEfficiency}. Each \ce{H+} driven into the cell by a PMF of $\approx \SI{150}{\milli\volt}$ \cite{Unden1998} can perform $w = 2.4 \times 10^{-20}\,\si{\joule}$ of work. If we assume that flagella motors extract work with 100\% efficiency from the protons, then $P/w \lesssim 2 \times 10^4 \,\ce{H+}\,\si{\per\second}$ is needed to power the `effective motor'. In reality, of course, a somewhat higher flux would be needed, because the motor is not 100\% efficient in harnessing energy from the PMF. 

On the other hand, full oxidation of one glucose gives 12 NADH; each NADH can export up to \ce{8H+}, Thus, $\dot G/2 \approx 4\,\si{\atto\mol\per\minute\per\mbox{cell}} \equiv 4 \times 10^4 \, \si{\per\second\per\mbox{cell}}$ fully-oxidised glucose molecules export up to $384 \times 10^4 \ce{H+} \si{\per\second\per\mbox{cell}}$. This is in large excess of the $\sim 10^4 \ce{H+} \si{\per\second\per\mbox{cell}}$ we estimate as needed to power the $\approx 50\%$ increase in $\bar v$ upon addition of glucose. Thus, swimming accounts for only a small part of a cell's energy budget. 

\subsection{Other small molecules}

We have also studied swimming powered by acetate (\ce{C2}), glycerol and lactate (\ce{C3}), xylose (\ce{C5}), galactose (\ce{C6}), and maltose and lactose (\ce{C12}), which can be expected to generate different amounts of energy per molecule. Interestingly, however, the maximum increase in speed in each case is approximately the same as observed for glucose (about $50\%$, see Fig.~\ref{fig:glucose}(d) for L-lactate). This suggests that the flagellar motor speed saturates at high PMF (as observed for {\it Bacillus subtilis} \cite{Shioi1980}), and supports our conclusion that swimming uses only a small fraction of metabolic energy. At high enough concentration of these alternative carbon sources, we find that the post-\ce{O2}-exhaustion speed drops to zero (for lactate, cf.~\cite{Douarche}), rather than to some intermediate, constant level as observed for glucose. This is because \ecoli\, cannot ferment these substrates. 

\section{\textbf{\textit{E. coli}} swimming powered by light}
\label{sec:light}

\begin{figure}[t]
\begin{center}
\includegraphics[width=0.48\textwidth]{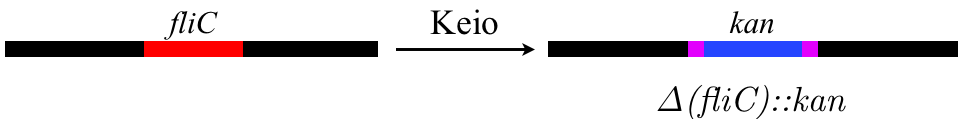}
\caption{A non-motile Keio knockout \cite{Keio}, $\mathit{\Delta}${\it (fliC)::kan}, where the flagella protein gene {\it fliC} (red) is replaced by the kanamycin resistance gene {\it kan} (blue), flanked by DNA sequences (purple) that allow its easy excision. }
\label{fig:keio}
\end{center}
\end{figure}

Some micro-organisms directly harvest energy from light. Thus, photosynthesizers use photons to split water to generate high energy electrons. Alternatively, light is harvested directly to pump protons using proton pumps such as proteorhodopsin (PR). PR, first discovered through the genomic sequencing of marine bacteria \cite{Beja2000}, has been incorporated into \ecoli\, \cite{Walter2007}.

Most PRs identified to date can generate PMF up to or slightly above WT levels \cite{Berry2013}. To enable illuminated PR to take over PMF generation, other sources need to be deactivated. This has been demonstrated on single cells using poisons for the respiratory enzymes (cf. Fig.~\ref{fig:PMF}) or by \ce{O2} exclusion. As the motor, and therefore flagellum rotation speed, is proportional to the PMF, illuminating poisoned cells expressing PR leads to a controlled increase in flagellum rotation speed \cite{Walter2007}.

Here, we demonstrate this effect in a cell population using MG1655 transformed with plasmid pBAD-HisC-PR expressing PR when induced by arabinose (DM-4). We already know that \ce{O2} is exhausted after a period  when swimmers are sealed in capillaries, Fig.~\ref{fig:endo_speed}.  After our transformed cells have exhausted \ce{O2}, their swimming is wholly powered by PR-generated PMF. The initial increase in speed with incident intensity, Fig.~\ref{fig:LightControl}, saturates at high intensities, in agreement with previous measurements of the rotation rate of individual flagellar motors \cite{Walter2007}. This and other PR mutants therefore enable simple external control of their swimming speed, making it feasible to change the samples activity both in time and space. They are living analogues of light-driven synthetic swimmers \cite{PalacciCrystal,BechingerJanus}.

\section{\textbf{\textit{E. coli}} -- the genetic toolbox}
\label{sec:genetools}

A great attraction of \ecoli\, is the availability of many mutants and a versatile genetic `tool box'. Thus, e.g., the FliC flagella protein could be mutated to include a cysteine residue to make disulphide bonds with dyes, or the {\it cheY} chemotaxis gene could be deleted to give smooth (non-tumbling) swimmers.  We briefly introduce two `off the shelf' mutant libraries and two methods for  mutagenesis to illustrate principles and introduce terminology. Details can be found in the textbooks, e.g. \cite{Dale}.

\begin{figure}
\begin{center}
\includegraphics[width=0.48\textwidth]{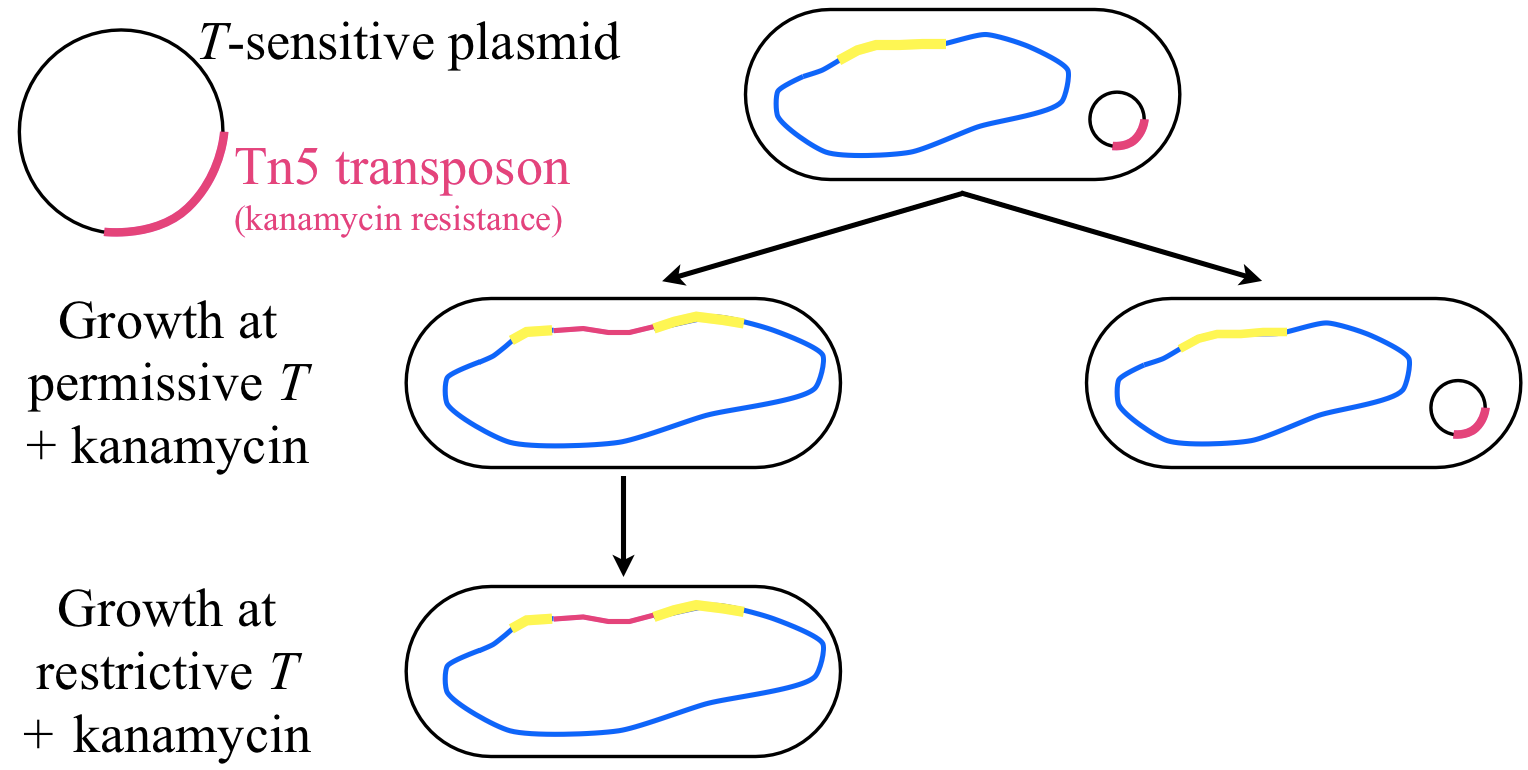}
\caption{Schematic of transposon mutagenesis. See text for details.
}
\label{fig:transmut}
\end{center}
\end{figure}

A major resource is the Keio collection \cite{Keio} of `single knockout' mutants of strain BW25113 (closely related to MG1655 but more amenable to molecular biology), available on a non-profit making basis to academics. Each  non-essential gene was precisely deleted and replaced with an excisable kanamycin resistance gene, Fig.~\ref{fig:keio}. 

Many mutagenetic methods use plasmids, extrachromosomal pieces of DNA that can be `transformed' into bacteria and maintained independently. Figure~\ref{fig:transmut} illustrates one strategy to carry out `transposon mutagenesis'. It relies on  a temperature-sensitive plasmid, which is only maintained within the cell at the `permissive' temperature, and is lost at a (higher) `restrictive' temperature.  This `suicide' plasmid acts as a `vector' (carrier) for a transposon, a piece of genetic code that can `jump' between DNA locations. Here, the plasmid carries transposon Tn5, which confers resistance to the antibiotic kanamycin.

When the transposon jumps, or transposes, from the plasmid onto the bacterial chromosome, it may insert into a gene,  generating a mutation.    Following a shift of the \ecoli\, from the permissive to the restrictive temperature, the plasmid will be lost. Then, all kanamycin-resistant cells carry a chromosomal copy of the Tn5 element transposed onto the chromosome, where it can be stably maintained.   Crucially, the transposon jumps into random chromosomal locations, and so can generate mutations in any gene.    A particular mutation can be selected by a change in phenotype and the   insertion site mapped by sequencing. 

P1 transduction is another popular mutagenesis technique.  Here DNA is transferred from a donor to a recipient cell using the P1 bacterial virus (a bacteriophage).   When P1 infects a host cell, it not only packages its own genome into the virus particle, but also can package fragments of the host chromosome.  When these virus particles are purified and used to infect the target strain of \ecoli, DNA homology between the donor DNA packaged in the virus particle and the recipient chromosome can generate insertion of the donor DNA.  

More generally, plasmids constitute a vital resource for introducing new genes into \ecoli, e.g. transforming with a plasmid carrying the GFP gene will render cells fluorescent. Each member in another useful library, the ASKA collection \cite{ASKA}, consists of an \ecoli\, strain containing a plasmid in which a single gene from \ecoli\, strain W3110 has been cloned.  The DNA encoding each gene has been fused to sequences that permit precise temporal control of the expression of the associated protein and its fluorescent tagging with GFP.  

We end with two `health warnings'. First, genetic modification in motility-related work typically starts with parent strains selected for motility (e.g. HCBxx or RPxx, Table~\ref{tab:strains}). Many of these retain wild-type features not optimised for DNA manipulation. Moreover, the genome of many `motility favourites' remains less well characterised compared to `canonical' strains such as MG1655 \cite{coligenome}. Sequencing and annotating the genome of some of these strains is a desirable future step. 

Secondly, care must be taken in obtaining a good reliable starter culture of the strain that one chooses to work with, either from the originating author  or from a national culture collection, e.g.~ATCC \cite{ATCC} and CGSC \cite{CGSC} in America; NCIMB \cite{NCIMB} in the UK; DSMZ \cite{DSMZ} in Germany; and CGMCC in China \cite{CGMCC}.  Bacterial genomes are changeable, so that strains deposited at the back of freezers twenty years ago whose provenance  is uncertain are not good starting points.

\begin{figure}[t]
\begin{center}
\includegraphics[width=0.45\textwidth]{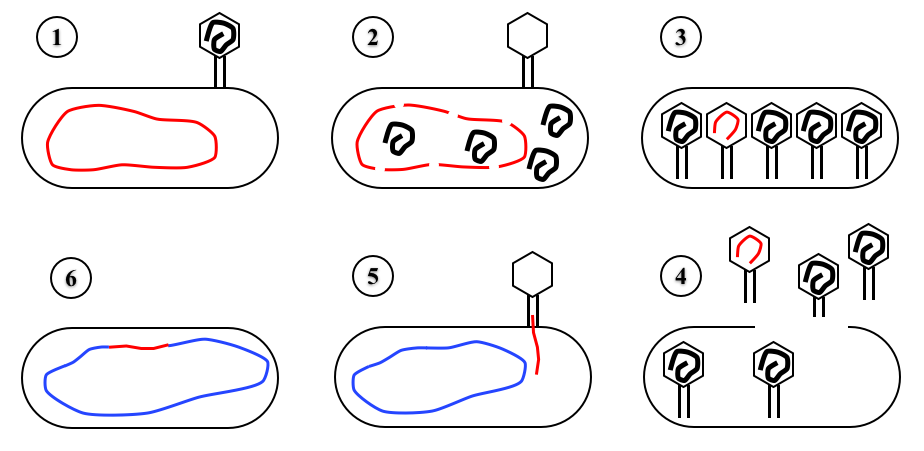}
\caption{Schematic of P1 phage transduction. \protect\circled{1} Phage attaches to cell. \protect\circled{2} Phage injects DNA (black) into cell and uses its machinery to replicate as well as fragment cell's genome (red). \protect\circled{3} DNA is packaged into viral particles, some of which contain fragments of the cell's DNA, giving a `Trojan' phage. \protect\circled{4} Cell lyses, releasing all viral particles. \protect\circled{5} `Trojan' phage infects target cell.  \protect\circled{6} Donor DNA fragment is incorporated into target cell genome by recombination. }
\label{fig:p1trans}
\end{center}
\end{figure}

\section{Summary and conclusions}

{\it Escherichia coli} is increasingly used in active colloids experiments as an alternative to or to contrast with synthetic colloidal swimmers. We have reviewed, explained and proposed a set of procedures that, together, should enable the collection of  data from such experiments between different laboratories that can be quantitatively compared. At the heart of these procedures is the use of a new, high-throughput method to characterise \ecoli\, motility, differential dynamic microscopy (DDM), which delivers the average swimming speed and motile fraction averaged over $\approx 10^4$ cells in a matter of minutes \cite{WilsonDDM,MartinezDDM}. Using DDM, we have established that \ecoli\, swims at constant speed when motility is powered by mixed-acid fermentation, Fig.~\ref{fig:glucose}(b). Along the way, we have explored the use of DDM to interrogate aspects of cellular physiology {\it via} the dependence of swimming speed on PMF. Our results suggest that \ecoli\, cells powered by endogenous metabolism actively reduce their PMF with time. It appears that about 50\% of the glucose under our conditions is used for generating energy, and that a very small fraction (maybe $\approx 1\%$) of this energy is used for motility.

Whether our proposed procedures, or something similar, are followed, it is important that sufficient information is recorded in publications using \ecoli\, as a model active colloid to enable faithful reproduction of experiments. We therefore propose a `checklist' of what needs to be reported in \ref{app:checklist}. 

More needs to be done before \ecoli\, can become an ideal model for active matter experiments. Biologically, it would be desirable to sequence the genomes of some favourite motility strains. Armed with accurate genomic information, it should be possible to design bespoke swimmers that have ideal characteristics for different aspects of active matter research, e.g., a strain that does not depend on either poison, or \ce{O2} depletion, or light to display constant swimming speed over extended time. Theoretically, it would be useful to construct  coarse-grained models that take into account the flagella bundle. Predictions from such theoretical models confronted with data from experiments using increasingly well-characterised \ecoli\, suspensions should propel the subject of active colloids significantly forward.

\section*{Acknowledgements} WCKP initiated work with \ecoli\, under an EPSRC Senior Research Fellowship (EP/D071070). He and JSL, JA, AD and VAM are now funded by an EPSRC Programme Grant (EP/J007404/1) and an ERC Advanced Grant (ADG-PHYAPS). VAM's was also supported by Marie Curie Fellowship `ActiveDynamics' FP7-PEOPLE (PIIF-GA-2010-276190). AJ held an EPSRC studentship. TV is funded through Marie Curie fellowship `Living Patchy Colloids' (LivPaC, 623364) under the FP7-PEOPLE-2013- IEF program. TP is supported by an Edinburgh Chancellor's Fellowship and the BBSRC/EPSRC/MRC Synthetic Biology Research Centre grant (BB/M018040/1).  

Many past lab members contributed to our knowledge of \ecoli\, motility, especially Otti Croze (who cultured our first \ecoli) and Laurence Wilson (who performed the first DDM). We thank Howard Berg (Harvard), Ian Booth (Aberdeen) and Gail Ferguson (Edinburgh, then Aberdeen) for educational and illuminating discussions, and all the authors represented in Table~\ref{tab:strains} for providing strains and plasmids. 

\appendix

\section{Protocol reporting checklist} 
\label{app:checklist}

To facilitate reproducing and comparing results, we suggest that published methods should always include:
\begin{itemize}
\item Strain name and details of any genetic modifications 
\item Growth conditions (temperature, growth medium composition, cell density of harvest) 
\item Washing procedure (method: filtration or centrifugation, number of washing steps, exact composition of motility medium) 
\item Conditions for motility experiments (cell concentration and how determined, sample cell details: material and dimensions, sealing, position and conditions for image/video acquisition) 
\item Details for image analysis and data processing
\end{itemize}

\section{Strains}
\label{app:strains}

Table~\ref{tab:strains} summarises a number of strains of \ecoli\, that have been used by us and others in active colloids and related motility work. Note that publication in a journal carries the obligation of curating and sharing the strains used in a published work.However, one should not give away a strain obtained from another laboratory; instead the protocol is to direct enquiries to the strain originator. Note that intra-strain variability can arise rapidly because certain genetic elements (called insertion sequences, IS) can `jump' into regions of the genome controlling flagella gene expression, enhancing the motility \cite{Barker2004}.

\begin{table*}[!h]
\caption{\ecoli\, strains and plasmids}
\label{tab:strains}
\begin{center}
\begin{tabular}{|l|l|l|l|l|}
\hline
Name  & Motility genotype/relevant characteristics &  Source \\ \hline \hline
\ecoli\, strains & & \\
\hline
MG1655 &  motility WT  &  Laboratory strain~\cite{coligenome}  \\ \hline
AB1157 & motility WT & Laboratory strain~\cite{Bachmann1972} \\ \hline
RP437 & motility WT & J. Parkinson \cite{ParkinsonTumbly} \\ \hline
HCB1 & motility WT & H. C. Berg \cite{WolfeBerg}\\ \hline
BW25113 &  Parent strain of Keio collection, motile  & \cite{Keio} \\ \hline
HCB437 &   \begin{tabular}[c]{@{}l@{}}$\mathit{\Delta}${\it (tsr)}7021 $\mathit{\Delta}${\it (trg)}100 $\mathit{\Delta}${\it (cheA-cheZ)}2209\\ Defective in chemotaxis, a smooth swimmer\end{tabular}  & H. C. Berg \\ \hline
DM4 & MG1655 + proteorhodopsin-expressing pBAD-HisC-PR & This work \\ \hline
JSL1 &  \begin{tabular}[c]{@{}l@{}} AB1157 $\mathit{\Delta}${\it (cheY::kan)} \\ Smooth swimmer \end{tabular}  & This work \\ \hline
AD1 & \begin{tabular}[c]{@{}l@{}} AB1157 {\it fliC}(S353C) \\ Motility WT with FliC mutation to bind to Alexa dye \end{tabular} &  This work \\ \hline
AD2 &  \begin{tabular}[c]{@{}l@{}} AD1 $\mathit{\Delta}${\it (cheY::kan)} \\ Smooth swimmer with FliC mutation to bind to Alexa dye  \end{tabular} &  This work \\ \hline
AD3 &  \begin{tabular}[c]{@{}l@{}} AD1 $\mathit{\Delta}${\it (ycgR::kan)} \\ Motility WT, unable to `brake'  \end{tabular}  & This work \\ \hline \hline
Plasmids & & \\
\hline
pHC60 & Tet$^{\rm R}$, GFP constitutive & \cite{Cheng1998} \\ 
\hline
pBAD-HisC-PR & Amp$^{\rm R}$, arabinose inducible proteorhodopsin & \cite{Berry2013} \\
\hline
\end{tabular}
\end{center}
\end{table*}

\section{Nomenclature for bacterial genetics}
\label{app:genotype}

Researchers working with bacteria, especially if they collaborate with biologists, need some facility in the shorthand used to described genotypes~\cite{adelberg,Maloy}, which in turn requires some knowledge of molecular genetics~\cite{Dale}. Here we give a very brief introduction based on `worked examples' from Table~\ref{tab:strains}. 

Each gene is given a three-letter, lower-case italicised name (possibly followed by an extra capital letter for different members of a gene family), often suggestive of its function; different versions of a gene (alleles) are distinguished by numbers. Thus, {\it tsr}7021 is allele 7021 of the {\it tar} gene in \ecoli\, coding for part of a receptor system for sensing L-serine and related amino acids.  Deletion of this allele in HCB427 is indicated by $\mathit{\Delta}${\it (tar)}7021. The protein product of a gene is denoted by the same three letters unitalicised and capitalised. Thus, in the {\it fliC}(S353C) mutation carried by AD1, residue 353 of the FliC protein used to build flagella filaments has been mutated from serine (S) to cysteine (C).\footnote{See the textbooks, e.g. \cite{Stryer}, for single-letter abbreviations of amino acids.}  Sometimes, a different gene is inserted in place of a deleted gene. Thus, in JSL1, a cassette of genes conferring resistance to the antibiotic kanamycin has been inserted in place of the deleted {\it cheY} gene for chemotaxis -- $\mathit{\Delta}${\it (cheY::kan)}; the advantage is that such mutants can be easily selected by growth in the presence of kanamycin. Finally, many bacteria carry plasmids, such as plasmid pHC60, which carries tetracycline resistance (Tet$^{\rm R}$) and expresses green fluorescence protein (GFP) constitutively (i.e. continuously rather than subject to environmental control). 

Genotype of strains can be found from various online databases, including the various culture collections \cite{ATCC,CGSC,NCIMB,DSMZ,CGMCC}. Information on individual genes are available from appropriate databases, e.g. \cite{EcoCyc}.

\section{Filtration}
\label{app:filter}

To transfer the bacteria from TB growth media to BMB, we use a sterile Nalgene filtration unit consisting of two compartments separated by a Milipore \SI{45}{\micro\meter} HATF filter \cite{BergTurnerDiff}. The filter is soaked in BMB for $\approx \SI{5}{\minute}$ before being placed centrally on the unit with sterile tweezers. After screwing tight the upper and lower halves of the unit, \SI{35}{\milli\litre} of cell culture is slowly poured onto the filter. A tap-powered suction pump is used to enhance the flow rate, with a 70\% Ethanol bath between the unit and the suction tap to prevent water supply contamination.

When the level of TB had fallen to $\approx \SI{3}{\milli\meter}$ above the filter, leaving $\approx \SI{3}{\milli\litre}$ of liquid, \SI{35}{\milli\litre} of BMB is pipetted into the unit. Ensuring that the filter does not run dry during the procedure minimises the number of non-motile cells. This washing step is repeated 3 times. The water flow is adjusted to maintain a constant filtration rate so that each wash step takes \SI{10}{\minute}. When $\sim$ 1 to 3~\si{\milli\litre} of filtrate remains after the third wash, a sterile cut pipette tip is used to transfer the filtrate to a plastic 50ml polystyrene test tube (Greiner). (Smaller tubes are not wide enough for next step.) The filter is removed from the unit using sterile tweezers and deposited on the side of the tube. By gently rolling the liquid over the filter, bacteria were resuspended to reach a final OD between 5 and 15, depending on the volume of liquid left on-top of the filter after the final washing step.

\section{Glucose assay}
\label{app:GluAssay}

We used a Sigma GAGO20 glucose assay kit. Hydrogen peroxide from the  oxidation of glucose by glucose oxidase reacts with o-dianisidine in the presence of peroxidase to form a colored intermediate, which further reacts with sulfuric acid to form a more stable, pink final product. The OD measured at 540 nm is proportional to the original glucose concentration. Calibration runs showed that in our equipment, the linear range for OD$_{\rm 540}$ {\it vs} glucose concentration extended to $\approx \SI{600}{\micro\mbox{M}}$.

\begin{table}[t]
	\begin{center}
	\caption{Glucose utilisation rate}
	\smallskip
	\begin{tabular}{|c| c|}
	\hline
	Initial [Glu] (mM) 	& rate ($\si{\micro\mole\per\minute\per\gram\,\mbox{wet cells}}$)	\\
	\hline
	0.50	& 4.9 $\pm$ 1.0	\\
	\hline
	0.20	& 5.2 $\pm$ 0.9	\\
	\hline
	0.10	& 6.1 $\pm$ 1.6	\\
	\hline
	0.05	& 4.7 $\pm$ 0.8	\\
	\hline
	\end{tabular}
	\end{center}
	\label{tab:util}
\end{table}

Bacterial cells were grown using the standard motility protocol, washed and OD adjusted to 0.3 as for motility experiments, except that the buffer here lacks EDTA, which interfered with glucose oxidase. Cells and glucose were mixed and samples taken at different time points, put on ice and filtered (\SI{0.22}{\micro\meter} pore size) as quickly as possible. The filtered supernatant was used according to manufacturer's protocol to measure glucose concentration. Glucose utilization rates were calculated using glucose and time difference as well as wet or dry weight of cells in 1ml of OD = 0.3 suspension. Results are reported as averages and standard deviations using different time periods measured in one set of experiments. There is no systematic trend. We therefore average the value determined at different [Glu] to give an estimated glucose utilisation rate of $5.2\pm0.6~\si{\micro\mole\per\minute\per\gram\,\mbox{wet cells}}$, or $\approx 5 \pm 0.6\, \si{\atto\mol\per\minute\per\mbox{cell}}$.


\section{List of symbols}
\label{app:symbols} 
\begin{table}[h]
\begin{center}
\begin{tabular}{|l|l|l|}
\hline
Symbol & Typical units & Brief definition \\
\hline \hline
$[ \ldots ]$ & \si{\micro\mbox{M}} or \si{\milli\mbox{M}} & concentration of \ldots \\
\hline
$\bar x$ or $\langle x \rangle$ & -- & average of $x$\\
\hline\hline
{\bf Latin} & &\\
\hline
$c$ & \si{\micro\mbox{M}} or \si{\milli\mbox{M}} & small molecule conc.\\
\hline
$D$ & \si{\micro\meter\squared\per\second} & diffusion coefficient\\
\hline
$d$ & \si{\milli\meter} & colony diameter \\
\hline
$\dot G$ & \si{\atto\mol\per\minute\per\mbox{cell}} & glucose consumption rate\\
\hline
$g$ & \si{\meter\per\second\squared} & gravitational acceleration\\
\hline
$L$ & \si{\micro\meter} & cell body + flagellum length\\
\hline
$l$ & \si{\micro\meter} & cell body length \\
\hline
$\ell_g$ & \si{\micro\meter} & gravitational length\\
\hline
$n$ & \si{\mbox{cells}\per\milli\litre} & cell density \\
\hline
$n$ & -- & refractive index\\
\hline
$Q$ & $\si{\atto\mole\per\minute\per\mbox{cell}}$& \ce{O2} consumption rate \\
\hline
$q$ & \si{\per\micro\meter} & scattering vector\\
\hline
$T$ & \si{\kelvin} or \si{\celsius} & temperature\\
\hline
$t$ & \si{\second} or \si{\minute} & time \\
\hline
$t_c$ & \si{\second} or \si{\minute} & time of speed crashes\\
\hline
$u(t)$& -- & $\bar{v}(t)/\bar{v}_0$\\
\hline
$V$ & \si{\cubic\micro\meter} & volume \\
\hline
$\bar{v}$ & \si{\micro\meter\per\second} & average speed \\
\hline
$\bar{v}_0$ & \si{\micro\meter\per\second}  & $\bar{v}(t=0) \equiv \bar{v}(0)$\\
\hline
$v_s$ & \si{\micro\meter\per\second} & sedimentation speed\\
\hline
$w_0(c)$ & -- & $w_0(c)=\bar{v}_0(c)/\bar{v}_0 (c=0)$\\
\hline\hline
{\bf Greek} & &\\
\hline
$\alpha$ & \si{\per\second} or \si{\per\minute} & growth rate\\
\hline
$\beta$ & -- & non-motile fraction \\
\hline
$\gamma$ & -- & $l/\sigma$\\
\hline
$\eta$ & \si{\milli\pascal\second} & viscosity \\
\hline
$\lambda$ & \si{\micro\meter} & wavelength {\it in vacuo} \\
\hline
$\mu$ & \si{\micro\meter\centi\meter\per\volt\per\second} & electrophoretic motility\\
\hline
$\nu$ & -- & number of protons (\ce{H+})\\
\hline
$\xi$ & \si{\pico\newton\second\per\micro\meter} & friction coefficient\\
\hline
$\rho$ & \si{\gram\per\cubic\centi\metre} & (mass) density\\
\hline
$\sigma$ & \si{\micro\meter} & cell body diameter\\
\hline
$\tau$ & \si{\second}& delay time, run time, life time \\
\hline
$\phi$ & -- & cell body volume fraction\\
\hline
\end{tabular}
\end{center}
\label{default}
\end{table}%




\bibliographystyle{model1-num-names}
\bibliography{active}







\end{document}